\documentclass[twocolumn,superscriptaddress]{revtex4}
\usepackage{graphicx}
\usepackage{epstopdf}
\usepackage{amsmath}
\def\BS#1{{\bf #1}}
\begin{document}
%\draft
%twocolumn
%[\hsize\textwidth\columnwidth\hsize\csname@twocolumnfalse\endcsname
\draft
\title{Dynamics in the presence of attractive patchy interactions}

 \author{Cristiano De Michele} \affiliation{ {Dipartimento di Fisica and
  INFM-CRS-SOFT, Universit\`a di Roma {\em La Sapienza}, P.le A. Moro
  2, 00185 Roma, Italy} }
 \author{Simone Gabrielli} \affiliation{ {Dipartimento di Fisica and
  INFM-CRS-SOFT, Universit\`a di Roma {\em La Sapienza}, P.le A. Moro
 2, 00185 Roma, Italy} }
\author{Francesco Sciortino} \affiliation{ {Dipartimento di Fisica and
  INFM-CRS-SOFT, Universit\`a di Roma {\em La Sapienza}, P.le A. Moro
  2, 00185 Roma, Italy} } 
  \author{Piero Tartaglia} \affiliation{
  {Dipartimento di Fisica and INFM-CRS-SMC, Universit\`a di Roma {\em
  La Sapienza}, P.le A. Moro 2, 00185 Roma, Italy} }
        
\begin{abstract}
We report extensive monte-carlo and event-driven molecular dynamics simulations of a liquid composed by particles interacting via hard-sphere interactions complemented by four tetrahedrally coordinated short-range  attractive ("sticky") spots, a model introduced  several years ago by Kolafa and Nezbeda [J. Kolafa and I. Nezbeda, Mol. Phys. 161 {\bf 87} (1987)].  To access the dynamic properties of the model we introduce and implement a   new  event-driven molecular dynamics algorithm suited to study  the evolution of hard bodies  interacting, beside the repulsive hard-core, with a short-ranged inter-patch square well potential. We evaluate the thermodynamic properties of the  model in deep supercooled states, where the bond network is fully developed, providing evidence of density anomalies. We show that, differing from models of spherically symmetric interacting particles,  in a wide region of packing fractions $\phi$ the liquid can be super-cooled without encountering the gas-liquid spinodal.  In particular, we suggest that there is one optimal $\phi$ (not very different from the  hexagonal ice $\phi$) at which the bond tetrahedral network  fully develops.   We find evidence of the dynamic anomalies characterizing network forming liquids. Indeed, around the optimal network packing, dynamics fasten both on increasing and decreasing $\phi$. Finally we  locate the shape of the isodiffusivity lines in the $(\phi-T)$ plane and establish the shape of the dynamic arrest line in the phase diagram of the model.
Results are discussed in connection to colloidal dispersions of sticky particles  and 
gel forming proteins  and their ability to form dynamically arrested states.
\end{abstract}
%82.70.Dd (colloids) , 64.70.Pf (glass transition) , and 82.70.Gg (gels) 61.20.Ja Computer simulation of liquid structure

\maketitle
\section{Introduction}
This article presents a detailed numerical study of the thermodynamics and of the dynamics of a model introduced  several years ago by Kolafa and Nezbeda\cite{Kol87a} as  a primitive model for water (PMW).  The model envisions a water molecule as a hard sphere (HS) whose surface is decorated by four short ranged  "sticky" spots, arranged according to a tetrahedral geometry, two of which mimic the protons and two the lone-pairs.
Despite its original motivation, the Kolafa and Nezbeda model is representative of the larger class of particles interacting via localized and directional interactions, a class of systems which includes, besides network forming molecular systems, also proteins\cite{Lom99a,Sea99c,Ker03a}
and newly designed colloidal particles\cite{Man03a}. Indeed,   recent developments in colloidal science are starting to provide particles with specific directional interactions\cite{Yet03aNature}.  In the same way as sterically stabilized colloids have become the ideal experimental model for realizing the hard-sphere fluid, novel physical chemical techniques will soon make available to the community colloidal analogs of several molecular systems. A colloidal water is probably not far from being realized. 

Recent work\cite{Zac05a} has focused on the dynamics of colloidal particles interacting  with a restricted number of nearest neighbors. In Ref.~\cite{Zac05a, Mor05a} particles are interacting  via a limited-valency square well model\cite{Spe94aMP,Spe95aMP,Spe96aMP},  imposing a many body constraint   on the maximum number $n_{max}$ of bonded interactions. It has been found that  when $n_{max} < 6$, a significant shrinking of the liquid-gas (or colloidal rich-colloidal poor)  spinodal takes place. A window of packing fractions values opens up in which it is possible to reach very low temperature (and hence states with extremely long bond lifetime) without encountering phase separation. This  favors the establishment of a spanning network of long-living bonds, which in the colloidal community provides indication of gel formation but which, in the field of network forming liquids, would be rather classified as glass formation.   The study of the dynamics of the PMW provides a test of the $n_{max}=4$ results, in the absence of many-body interactions and in the presence of a geometric correlation between the bonding sites, retaining the maximum valency. This article, by reporting results on a model which can be at the same time considered as a simple model for the new generation of  patchy colloids  or for network forming liquids, starts to bridge the gap between these two fields.

 Thermodynamic and structural properties of several primitive models for water (and  other bonded systems) have been studied  in detail during the last 30 years\cite{Bra85a,Kol87a,Nez89a,Nez90a,Veg98a}, since this type of primitive models have become one of the landmarks for testing theories of 
association\cite{Wer84a,Wer84b,Gho93a,Sea96a,Dud98a,Pee03a,Kal03a}.  In particular, the theory of  Wertheim\cite{Wer84a,Wer84b}  has been carefully compared to early numerical studies, suggesting a good  agreement between theoretical predictions and  numerical data, in the temperature and packing fraction regions  where it was possible to achieve numerical equilibration\cite{Veg98a}. Recently, the increased numerical facilities, extending the range of studied state points, have clarified that deviations from the theoretical predictions start to take place as soon as the number of bonds (between different patches) per molecule increases and a network of bonded particles appears\cite{Vlc03a,Dud98a}. Geometric correlations between different bonds, not included in the theory, are responsible for the break down of the theoretical and numerical agreement.  Attempts to extend the perturbation theory beyond first order do not appear to be able to cure the problem\cite{Vlc03a}. The PMW is a good candidate for testing new theories of association and, for this reason,  it is important to clearly establish numerically the low $T$ behavior of the supercooled liquid state. The equilibrium PMW phase diagram, recently calculated\cite{Veg98a}, includes two crystal regions and a metastable fluid-gas coexistence.

All previous studies of primitive models for sticky directional interactions have focused on thermodynamic and static properties of the model. 
But the ability of fully exploiting the fast developments taking place in colloidal  physics
\cite{glotzer0,glotzer1} requires understanding not only the equilibrium phases of systems of patchy particles  and their modifications with the external fields, but also understanding the kinetic phase diagram\cite{Sci02a}, i.e. the regions in phase space where disordered arrested states can be expected, and when and how these states are kinetically stabilized with respect to the ordered lowest free energy phases.  In this respect, it is worth starting to establish the
dynamic properties of  simple models of patchy interactions, since the simplicity of these models (based on  hard sphere and square well interactions)  have the potentiality to provide us with an important  reference frame  and may play a relevant role in deepening our understanding of the dynamic arrest in network forming liquids, in connecting arrest phenomena associated to gel formation\cite{Zac05a,delgado} (the establishment of a percolating network of long lived bonds) and arrest related to excluded volume effects
and  the dependence of the general dynamic and thermodynamic features on the number and spatial location of patchy interactions.    
The case of the PMW reported here is  a good starting one. In this article we report thermodynamic data, extending the previously available information to lower temperatures and, for the first time, dynamic information obtained solving the Newton equations using a new algorithm based on event-driven propagation. 

\section{The Model and Numerical Details}
In the PMW, each particles is composed of an hard sphere of
diameter $\sigma$  (defining the length scale) and by four additional sites located along the direction of a tetrahedral geometry. Two of the sites (the proton sites H) are located on the surface of the hard sphere, i.e. at distance $0.5 \sigma$ from the center. The two remaining sites  (the lone-pair sites LP) are located at distance $0.45 \sigma$.
Besides the hard-sphere interaction, preventing different particles to sample distances smaller than $\sigma$, only the H and LP sites of distinct particles interact  via a square well (SW) potential $u_{SW}$
of width $\delta=0.15 \sigma$ and depth $u_0$, i.e. 
\begin{eqnarray}
u_{SW}=-u_0~~r<\delta \\ 
\nonumber
~~~~~=0~~~~~~r>\delta,
\end{eqnarray}
where $r$ is here the distance between H and LP sites.  The choice of $\delta=0.15 \sigma$ guarantees that multiple bonding can not take place at the same site.  The depth of the square well potential $u_0$ defines the energy scale.  Bonding between different particles is thus possible only for specific orientations and distances. In the linear geometry, the maximum center-to-center distance at which bonding is possible is  $1.1\sigma$ since the LP site is buried $0.05\sigma$ within the hard-core, a value typical of short-range colloid-colloid interactions. 

We have studied a system of $N=350$ particles with periodic boundary conditions  in a wide range of packing fraction $\phi \equiv \pi/6 n \sigma^3$ (where  $n$  is the number density)
and temperatures  $T$, where $T$ is measured in units of $u_0$ ($k_B=1$). We perform both Monte Carlo  (MC) and event driven 
molecular dynamics. In one MC step, an attempt to move 
each particle is performed.  A move is defined as a 
displacement  in each direction of a random quantity distributed uniformly between $\pm 0.05~\sigma$ and a rotation around a random axis of random angle distributed uniformly 
between $\pm 0.5$ radiant.   Equilibration was performed with MC, and monitored via the evolution of the potential energy
(a direct measure of the number of bonds in the system). 
The mean square displacement (MSD) was also calculated to guarantee that each particle has diffused in average more than its diameter. In evaluating the MSD we have taken care of subtracting the center of mass displacement, an important correction in the low $T$ long MC calculations. At low $T$ simulations required more than $10^9$ MC steps, corresponding to several months of CPU time.  

We have also performed event driven  (ED) molecular dynamic simulations of the same system, modeling particles  as constant density spheres of diameter $\sigma$ and mass $m$. The momentum of inertia 
is diagonal and equal to   $m \sigma^2/10$ . The algorithm implemented to propagate the newtonian trajectory in the presence of patchy square well interaction is described in details in Appendix~\ref{appendicecris}.   In ED dynamics, time is measured in units of $\sigma \sqrt{m/u_0}$.  Assuming as $m$ the mass of
the water molecule, as $u_0$ a typical value for hydrogen bond ($\approx 20 kJ/mol)$ and as $\sigma$ the nearest neighbor distance in water ($0.28 nm$), the unit of time corresponds  $\approx 0.3 ps$. All static quantities have been evaluated
with both MC and MD configurations finding no differences.   

Pressure, measured in units of $u_0/\sigma^3$, has been calculate as sum of three contributions. A trivial kinetic contribution, equal to $nk_BT$.
A positive HS contribution and a negative contribution arising from the SW interaction. Details of
the calculation of $P$ in both MC and ED simulations is provided in the Appendix \ref{pressure}.

\section{Results: Static}

\subsection{Potential Energy $E$}
Since in the PMW each site can take part to only one bond, due to geometric constraints fixed by the small value of $\delta$,
the lowest energy configuration is defined by four bonds per
particles, corresponding to a ground state 
energy per particle $E_{gs}=-2 $ (in units of $u_0$). Of course, this absolute ground state value may not be accessible at all $\phi$, due to the strong constraints introduced by the bonding geometry. 
According to Wertheim's  first order thermodynamic 
perturbation theory, the $T$ and $\phi$ dependence of
the potential energy per particle $E$  is given by\cite{Kol87a,Nez89a,Veg98a}
\begin{equation}
E-E_{gs}= \frac{2}{1+c}
\label{eq:u}
\end{equation}
where
\begin{equation}
c=0.5 \bigg\{  \left[  1+192 ( e^{\frac{1}{T}} -1) \phi J  \right]^{0.5} -1  \bigg\}
\end{equation}
\begin{equation}
J=\frac{c_1 (1-\phi/2)-c_2 \phi (1+\phi)}{(1-\phi)^3}
\end{equation}
with $c_1=2.375 \times 10^{-5}$ and $c_2=2.820 \time 10^{-6}$\cite{Veg98a}.
The Wertheim theory, which assumes uncorrelated
independent bonds, predicts  as low $T$ limit of Eq.~\ref{eq:u}  an Arrhenius $T$-dependence, 
\begin{equation}
\lim_{T \rightarrow 0} E-E_{gs}= \frac{4}{\sqrt{192 \phi J}} e^{-0.5/T}
\label{eq:ulowT}
\end{equation}
i.e. with an activation energy of half bond energy. It is worth observing that such an Arrhenius law, with an activation energy equal to $0.5$ characterizes the 
low $T$ dependence of the energy in the $n_{max}$ model\cite{Mor05a,Zac05a} 
[a model of particles interacting via a SW potential with an additional constraint on the maximum number of bonds], where no geometric correlation between bonds is imposed.

\begin{figure}[tbh]
\centering
\vspace{0.10cm}
\includegraphics[width=0.45\textwidth]{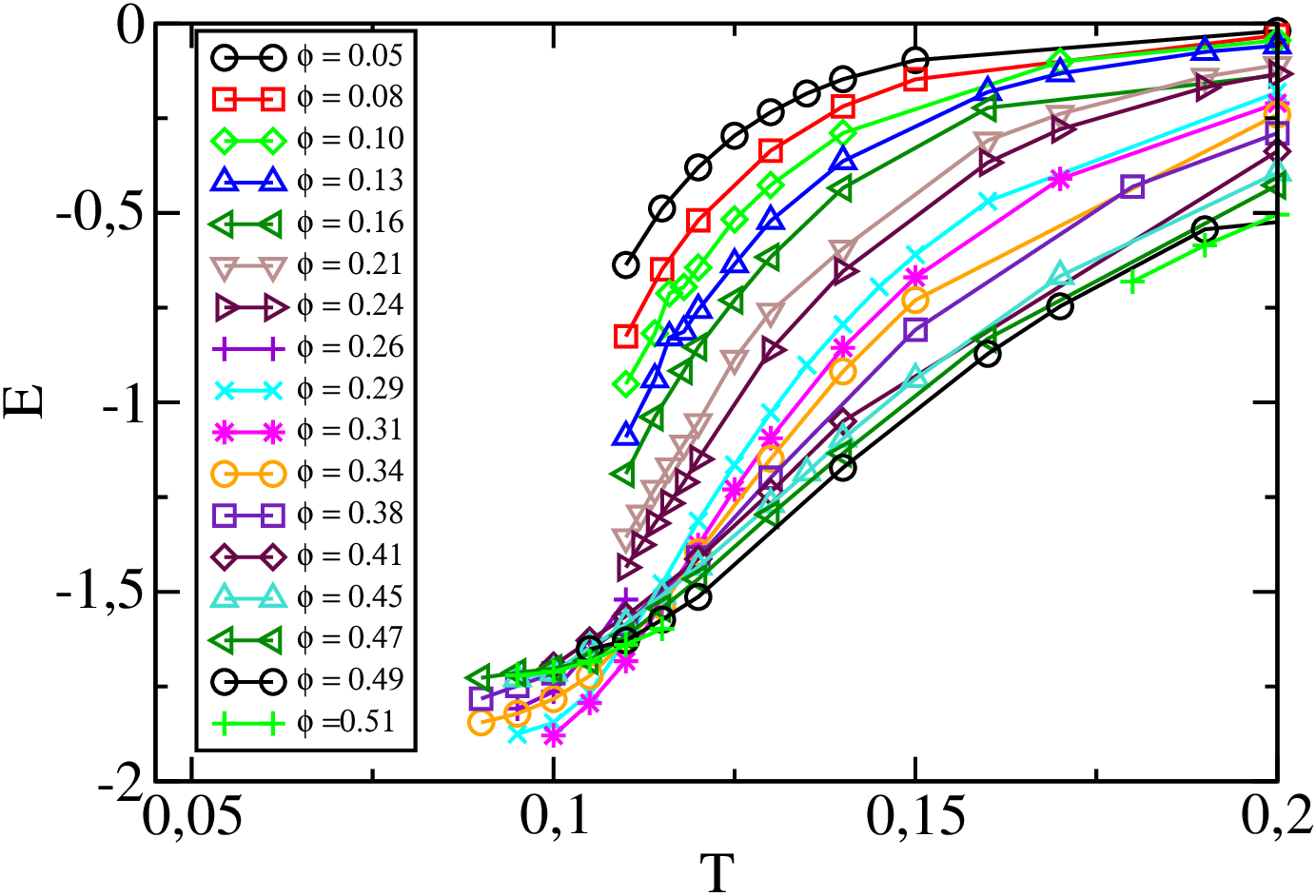}
\includegraphics[width=0.45\textwidth]{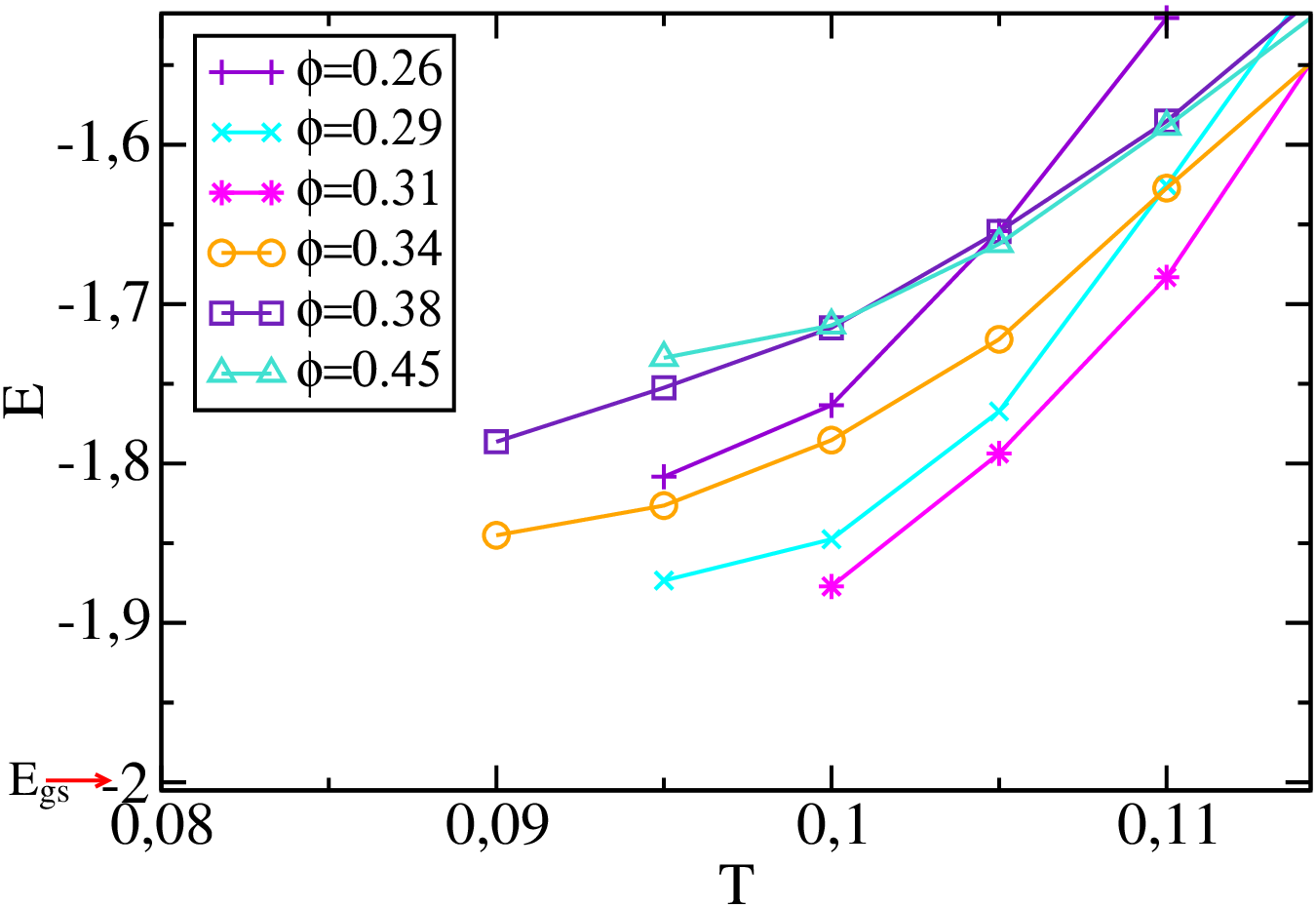}
%\vspace{0.10cm}
\caption{Potential energy for the PMW. The top panel shows data for all studied isochores as a function of $T$. The lower panel shows an enlargement of the low $T$ region, where the network is fully developed. Note that for this model, the lowest possible energy is $E_{gs}=-2$.}
\label{fig:ene}
\end{figure}

Fig.~\ref{fig:ene} shows the $T$ dependence of the potential energy  for different isochores.  As discussed in details in the following, for $\phi \lesssim 0.24$ a phase separation is encountered on cooling, preventing the possibility of equilibrating one-phase states below $T \approx 0.11$. For $\phi > 0.24$ the system remains homogeneous down to the lowest investigated $T$. The low $T$-behavior is expanded in Fig.~\ref{fig:ene}-(bottom). With the extremely long equilibration runs performed, proper equilibration is reached only for $T \gtrsim 0.09$. The enlargement of the low $T$ region shows that the absolute ground state  value $-2 u_0$ is closely approached at $\phi \approx 0.3$.  At higher or smaller $\phi$, the potential energy appear to approach a constant value larger than $-2 u_0$. 
Consistent with previous claims\cite{Veg98a}, high $T$ data are very well represented by first order thermodynamic perturbation theory. Systematic deviations  between theory and simulation data appears as soon as the number of bonds per particle becomes bigger than one. 
Comparing the simulation data with the Wertheim theory, it is confirmed that the
physics of the network formation is completely missing in the  first-order perturbation theory.

\begin{figure}[tbh]
\centering
%\vspace{0.10cm}
%\includegraphics[width=0.5\textwidth]{evsrho.eps}
\includegraphics[width=0.45\textwidth]{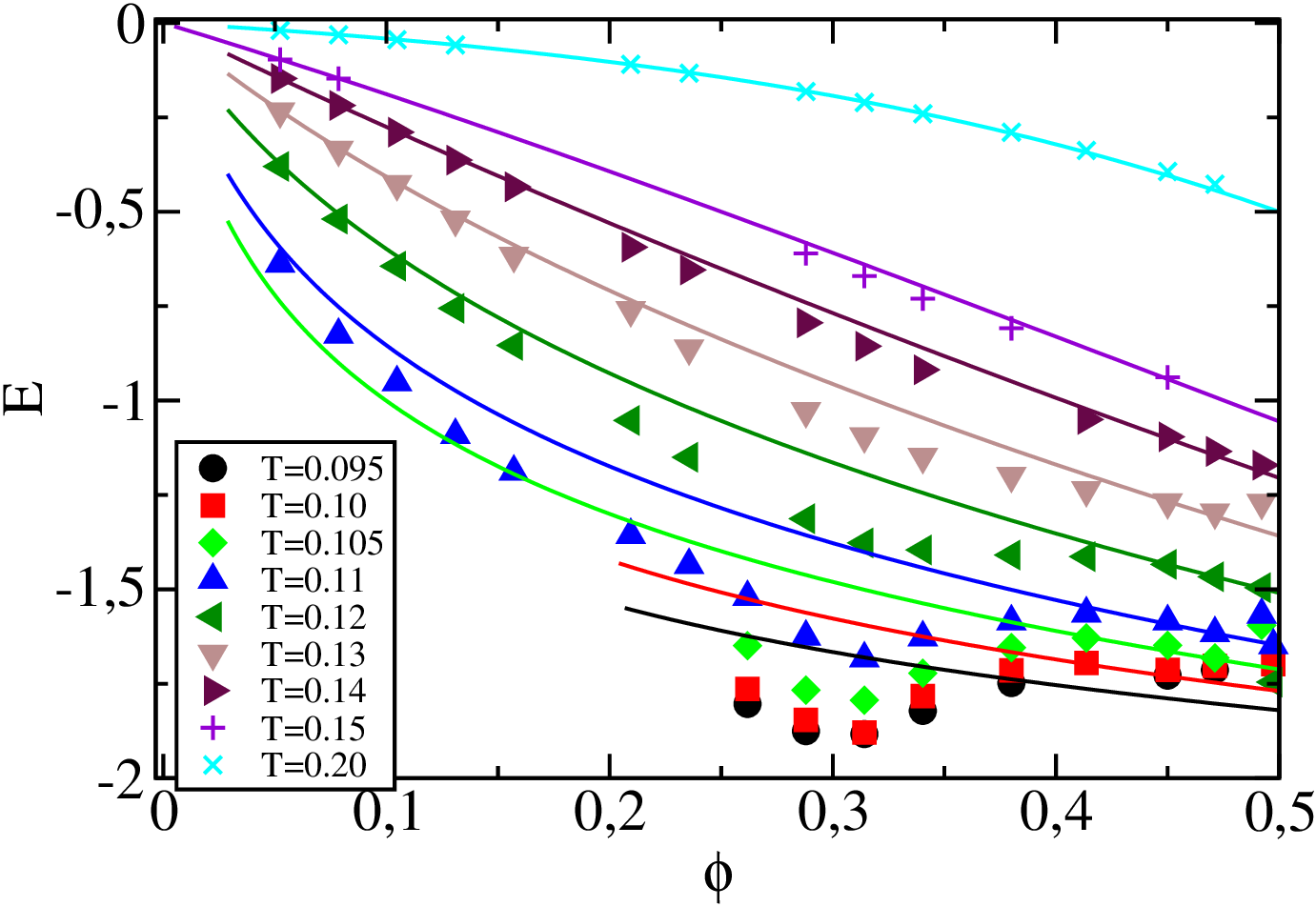}
%\vspace{0.10cm}
\caption{Potential energy vs. $\phi$ along isotherms. Symbols: simulation data. Lines: Wertheim's theory.}
\label{fig:enevsrho}
\end{figure}
Fig.~\ref{fig:enevsrho} shows the $\rho$ dependence of $E$ along
isochores.   At high $T$ ($T>0.13$), a monotonic decrease of $E$ is
observed, caused by the increased bonding probability
induced by packing.  In this $T$ region, the number of bonds is at most of the order of  two per particle. Completely different is the situation for lower $T$. The $\phi$ dependence becomes non-monotonic. There is a specific value of the packing fraction ($\phi \approx 0.3$) at which the lowest energy states are sampled.  In the following we define the optimal network packing fractions as the
range of packing fractions for which  it is possible to fully satisfy the bonds in a disordered homogeneous structure.   At  $\phi \approx 0.3$, the number of bonds at the lowest investigated $T$ (the lowest $T$ at which equilibration is feasible with several months of computation time) is about $3.8$ per particle, i.e. about 95\% of the bonds are satisfied.  The range of optimal $\phi$s appears to be rather
small. Indeed for packing fractions lower or higher than this optimal $\phi \approx 0.314$, the formation of a fully connected network is hampered by geometric constraints: at lower $\phi$, the large inter-particle distance   acts against the possibility of forming a fully connected network, while at large $\phi$, packing constraints, promoting close packing configurations are inconsistent with the tetrahedral bonding geometry.   Not surprisingly,
$\phi=0.314$ is within the range of $\phi$ values which allow for a stable open
diamond crystal phase ($0.255<\phi<0.34$)\cite{Veg98a}.  A reduction of the
geometric constraints (as in the $n_{max}$ model\cite{Zac05a,newzacca}) increases the
range of optimal $\phi$. It is worth also noting that
the liquid side of the spinodal curve is close to the region of optimal network $\phi$.

The existence of a convex form for the potential energy (here for $\phi \gtrsim 0.3$)  has been observed in several other models for tetrahedral networks, including models for water (and water itself\cite{Sci97c}). It has been pointed out that a negatively convex $\phi$ dependence is indicative of a  destabilization of the free energy\cite{Sci97c} and a precursor of a possible liquid-liquid critical point (in addition to the 
lower $\phi$ gas-liquid one). Liquid-liquid critical points have been observed in several 
models for water\cite{Poo92a,Poo93c,Yam02a,Poo05JPCM,Pas05aPRL,Bro05JCP}. Indeed, the Helmholtz free energy $A$ is related to $U$ (the sum of the kinetic and potential energy)  via $A=U-TS$, where $S$ is the entropy. The curvature of an isotherm of $A$ must be positive for a homogeneous phase of a specified volume $V$ to be thermodynamically stable.
The curvature of $A$ can be expressed as
\begin{equation}
\left( \frac{\partial^2 A}{\partial V^2}\right)_T=
\left( \frac{\partial^2 U}{\partial V^2}\right)_T-T
\left( \frac{\partial^2 S}{\partial V^2}\right)_T
\label{eq:d2a}
\end{equation}
Since $P=-\left( \frac{\partial A}{\partial V}\right)_T$
the inverse compressibility $K_T= -1/V (\partial V/\partial P)_T$ is related to the curvature of $A$
by
\begin{equation}
\frac{1}{ K_T}= V \left[ \left( 
\left( \frac{\partial^2 U}{\partial V^2}\right)_T-T
\left( \frac{\partial^2 S}{\partial V^2}\right)_T
\right) \right]
\end{equation}
The curvature of $A$ is thus proportional to $1/K_T$ for fixed
$V$. Since $1/K_T$ must be positive for a thermodynamically
stable state, for the range of $V$ in which $\left( \frac{\partial^2 U}{\partial V^2}\right)_T<0$, the contribution of the internal energy reduces the thermodynamic stability of the liquid phase. 
%This is confirmed by the fact that the range of V in which we find negative curvature in the $U-V$ data corresponds to the range
%in which the $K_T$ maxima are observed. 
The liquid remains stable where $U$ has negative curvature only because the contribution of the entropic term in Eq. ~\ref{eq:d2a} is large enough to dominate. Yet entropic contributions to these thermodynamic quantities are suppressed as $T$ decreases, due to the occurrence
of the factor of $T$ in the second term on the right-hand
side of Eq.~\ref{eq:d2a}. Hence the $U-V$ data suggest that at lower $T$ a single homogeneous phase of the
liquid will not be stable for certain values of $V$, leading to a
separation into two distinct liquid  phases of higher and lower volume.
Due to the predominant role of $E$ in the free energy at low $T$, the possibility of a phase separation of the PMW liquid into two liquid phases of different $\phi$, for  $\phi>0.3$ and $T$ lower than the one we are currently able to equilibrate should be considered.

\begin{figure}[tbh]
\centering
%\vspace{0.10cm}
\includegraphics[width=0.45\textwidth]{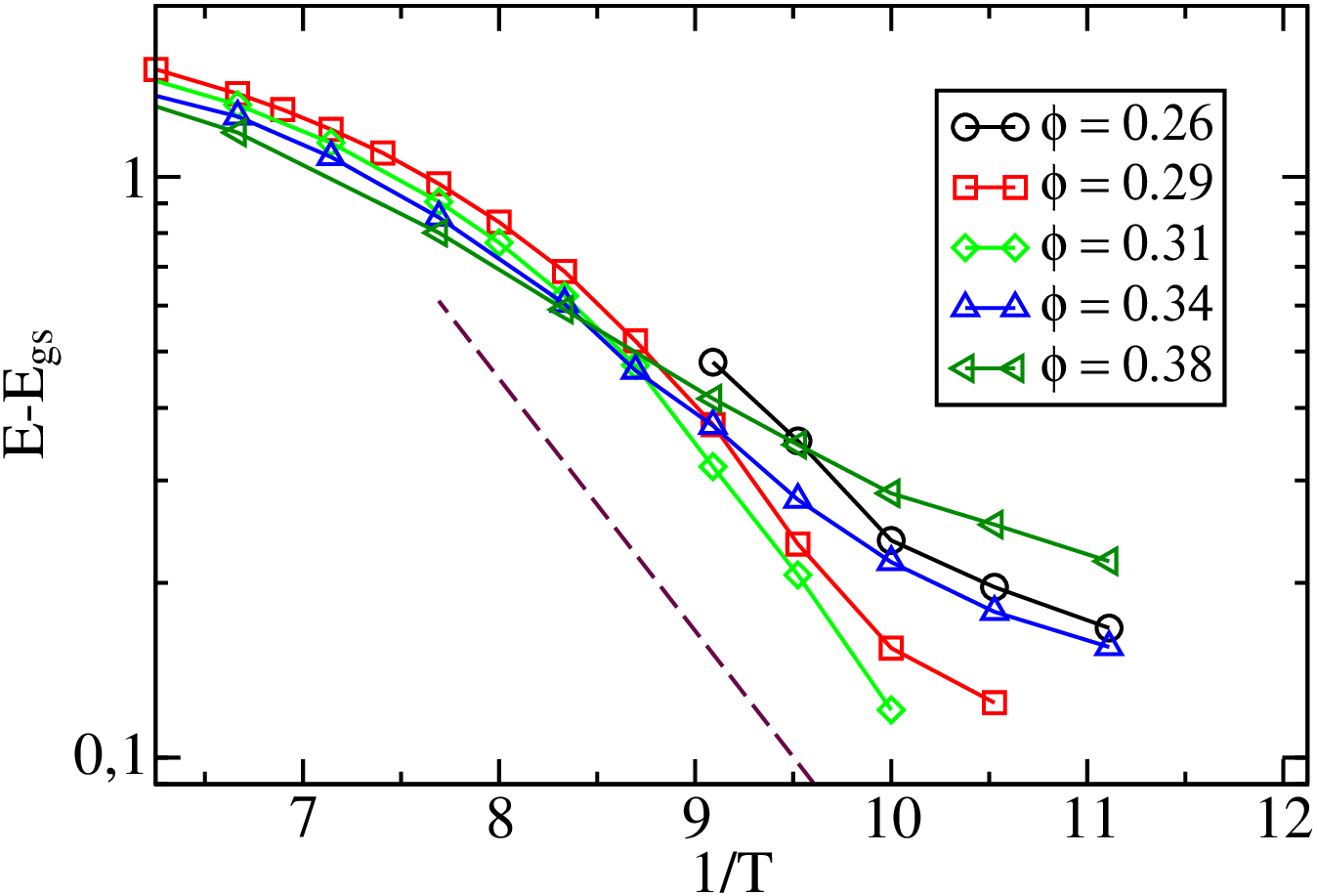}
%\vspace{0.10cm}
\caption{Arrhenius representation  ($E-E_{gs}$ vs $1/T$) of the potential energy around the optimal network density. The dashed line, shown as a reference, has an activation energy of $-u_0$.}
\label{fig:eneusut}
\end{figure}

Fig.~\ref{fig:eneusut} shows $\ln(E-E_{gs})$ vs $1/T$.  
 At the optimal $\phi$, the energy of the fully connected state is
 approached with an Arrhenius law,  characterized by an activation energy of $\approx 1 u_0$, clearly   different from the $0.5$ value predicted by the
 Wertheim theory.  For larger $\phi$ values, data  suggest that the lowest reachable state has an energy different from $-2 u_0$, consistent with the expectation that on increasing $\phi$, geometric constraints forbid the development of a fully connected network even at the lowest $T$. 

\subsection{$P$}
\begin{figure}[tbh]
\centering
\vspace{0.10cm}
\includegraphics[width=0.5\textwidth]{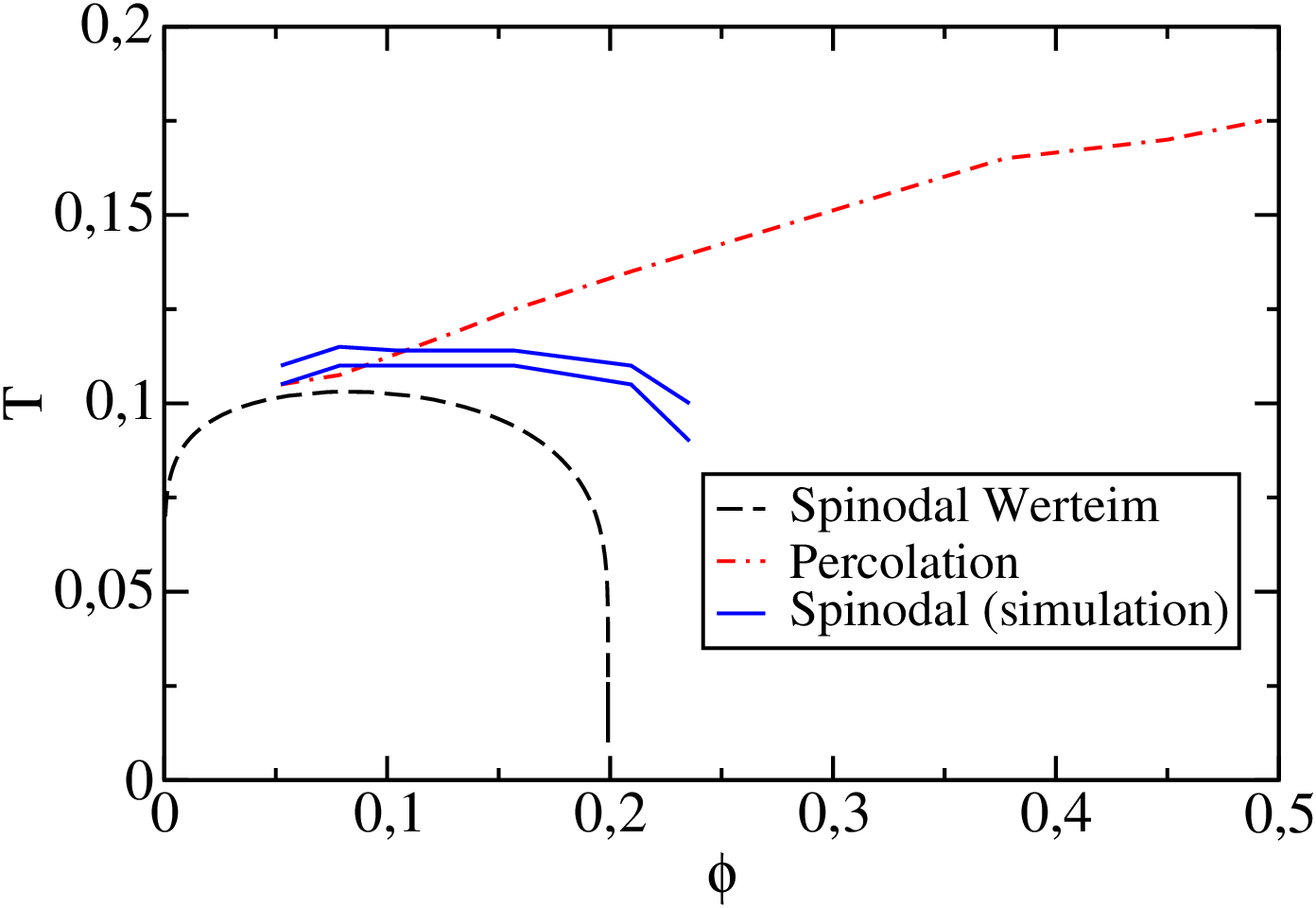}
\vspace{0.10cm}
\caption{Thermodynamic phase diagram for the PMW. The theoretical Wertheim prediction for 
the locus at which $(\partial P/\partial V)_T=0$ is compared with numerical estimates of the spinodal,
calculated by bracketing it via the locus of point at which $S(0) \approx 5$ and the
locus of points where a clear phase separation is detected.  The location of the bond percolation line
is also reported.}
\label{fig:phase}
\end{figure}

The Wertheim's prediction for the $T$ and $\phi$ dependence of the PMW pressure (the equation of state) is
\begin{eqnarray}
P=P_{HS}~~~~~ \\
\nonumber
- n k_BT  \frac{96 (e^{1/T} -1)}{(1+c)^2} 
\frac{c_1 \phi ( 1+ \phi - 0.5 \phi^2 ) - 2 c_2 \phi^2 (1+2 \phi)}{(1-\phi)^4}
\end{eqnarray}

where $P_{HS}$ is the pressure of the HS fluid at the same packing
fraction. $P_{HS}$ is very well represented by the 
Carnahan-Starling EOS\cite{Han86a} 
\begin{equation}
P_{HS}= n k_B T \frac{( 1 + \phi + \phi^2 - \phi^3)}{(1 - \phi)^3} 
\end{equation}
The Wertheim EOS predicts a vapor-liquid critical point 
at $T_c=0.1031$ and $\phi_c=0.085  $\cite{Veg98a}.  The vapor-liquid spinodals 
calculated according to the Wertheim theory and from simulation data 
are reported in Fig.~\ref{fig:phase}. The numerical estimate  is provided by
locating, along isochores, the highest $T$ state point in which
phase separation is observed and the $T$ at which the small $q$ limit of the structure factor 
is smaller than five.  These two state points bracket  the spinodal locus. 
It is interesting to compare the liquid-gas spinodal of the PMW with the corresponding spinodal of the symmetric spherical square well potential with same depth and well width $\delta=0.15$. In that case, the critical point is located at $T_c \approx 0.56$ and $\phi_c \approx 0.212$\cite{Pag05aJCP} and the high packing fraction (the liquid) side of the spinodal  extends beyond  $\phi=0.6$.  The net result of decreasing the surface available to bonding and of limiting to four the maximum number of nearest neighbors which can form bonds is the opening of a wide region of $\phi$ values where (in the absence of crystallization) an homogeneous
fluid phase is stable (or metastable). This finding is  in full agreement with the recent work of \cite{Zac05a}, where a  saturated square well model was studied for different values of the maximum valency. Indeed, it was found that when the number of bonds becomes less then six,  the unstable region  (the surface in the $(\phi-T)$ plane encompassed by the spinodal line) significantly shrinks, making it possible to access low $T$ states under single phase conditions. 

\begin{figure}[tbh]
\centering
\vspace{0.10cm}
\includegraphics[width=0.45\textwidth]{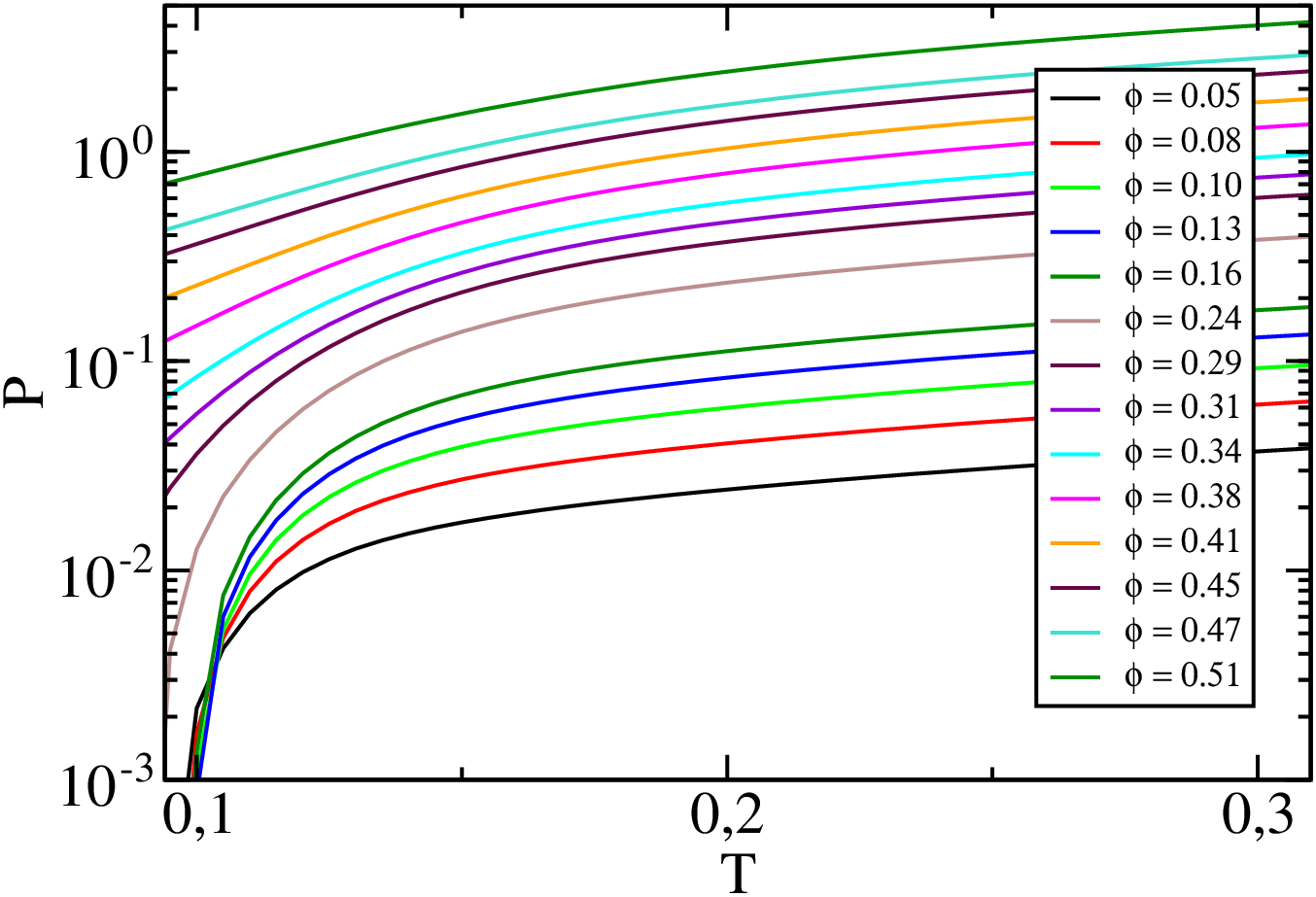}
\includegraphics[width=0.45\textwidth]{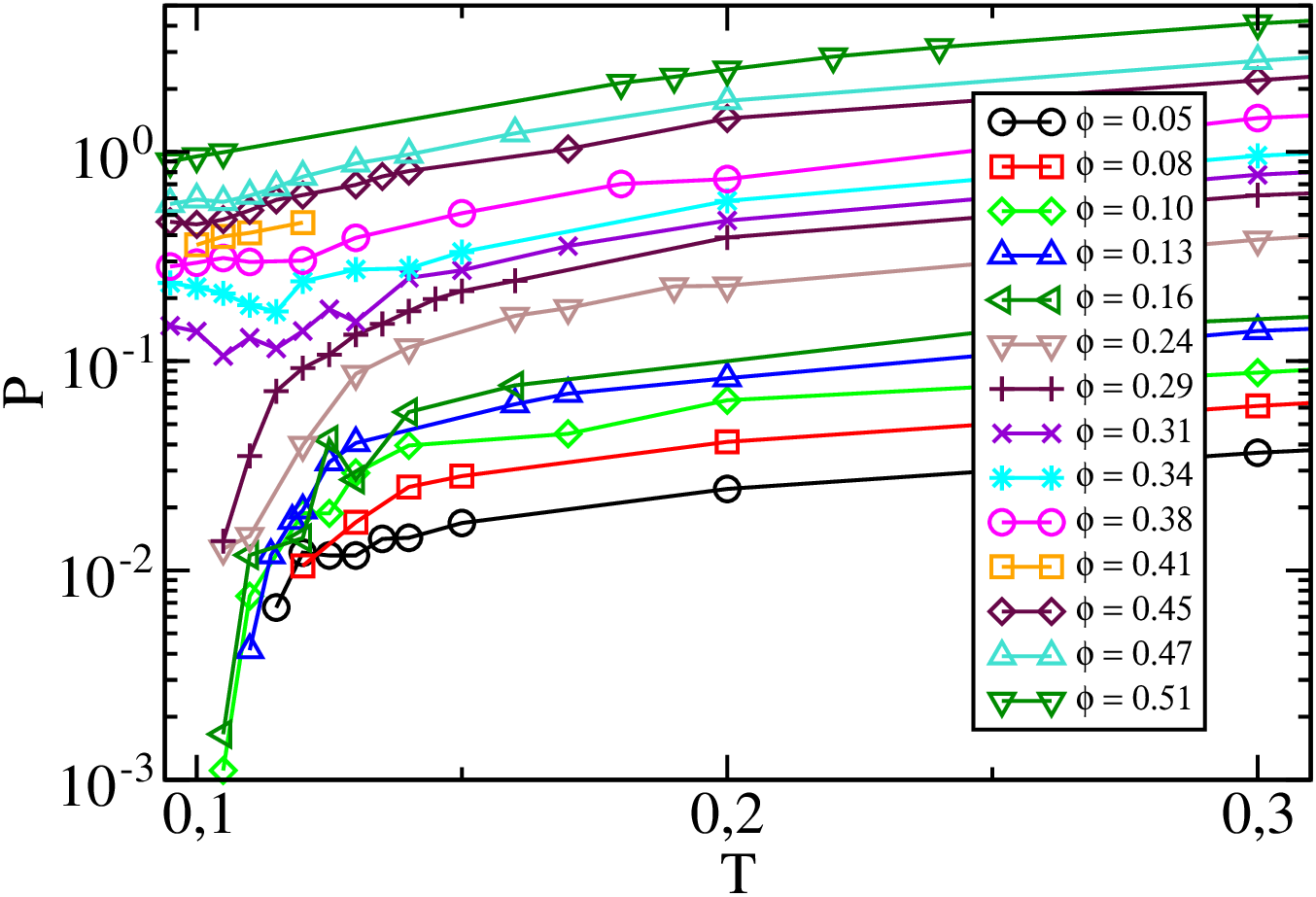}
%\vspace{0.10cm}
\caption{Isochores of $P$ according to the Wertheim theory (top) and as calculated from the simulation data (bottom).  Symbols refer to simulation data. The same sequence of $\phi$ values is shown in both panels.}
\label{fig:pvsT}
\end{figure}

\begin{figure}[tbh]
\centering
\vspace{0.10cm}
\includegraphics[width=0.45\textwidth]{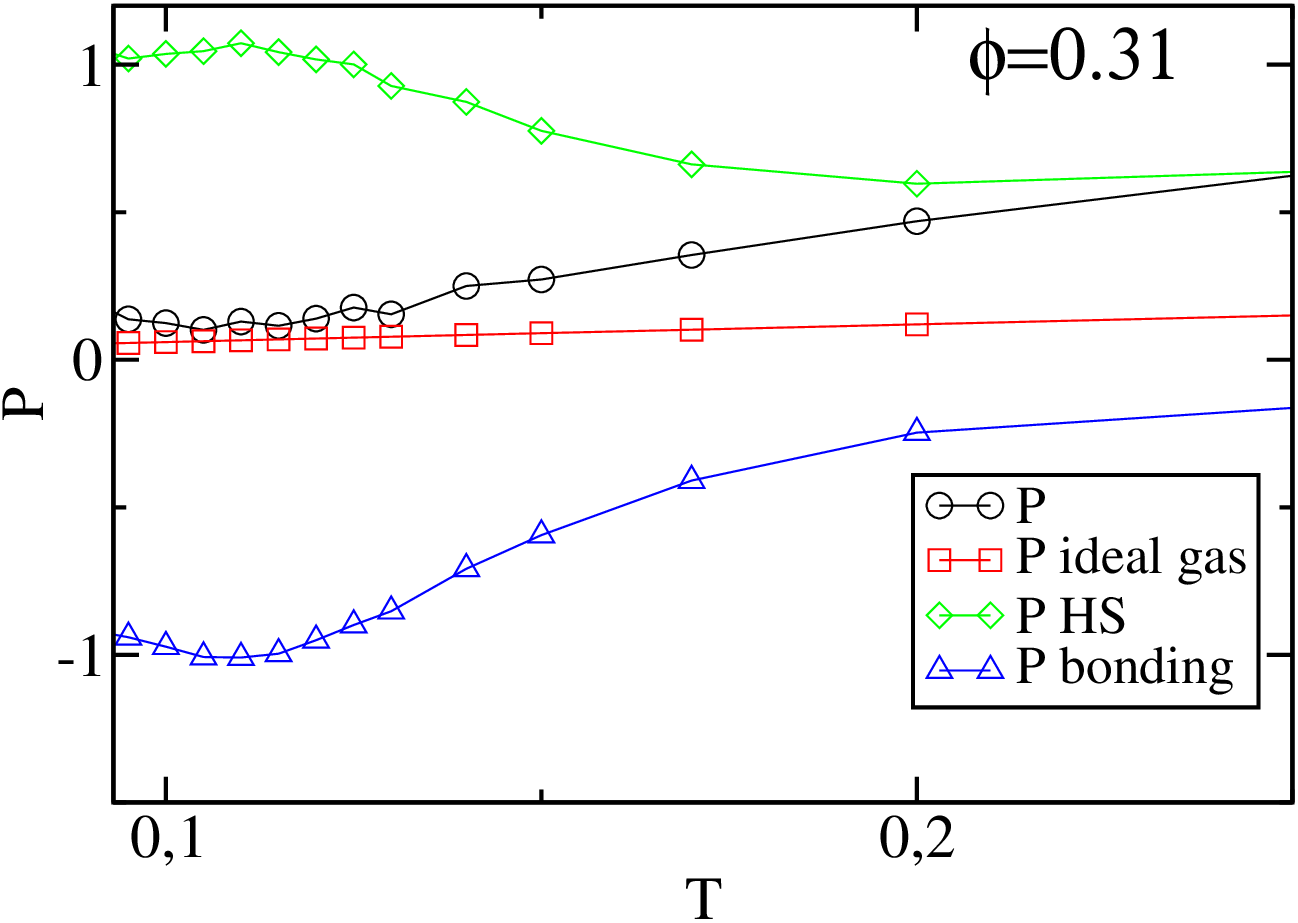}
%\vspace{0.10cm}
\caption{Components of the pressure at $\phi=0.314$. The total $P$ is decomposed in ideal gas, HS and bonding components. Note the isochoric minimum in $P$ around $T=0.105$, a signature of an isobaric density maximum.}
\label{fig:pcomp}
\end{figure}

Fig.~\ref{fig:pvsT} shows $P(T)$ for different isochores. In agreement with previous analysis, $P$ is well represented  by the Wertheim theory only at high temperature.    At low $T$ several interesting features are observed:
(i) for $\phi<0.25$, isochores end in the spinodal line.
(ii) in the simulation data, a clear difference in the low $T$ behavior is observed  between the two studied isochores $\phi =0.288$ and
$\phi=0.314$  . While in the $\phi =0.288$ case $P(T)$ decreases continuously on cooling, in the
$\phi=0.314$   case  the low $T$ behavior of $P$ is reverses and $P$ approaches 
a positive finite value on cooling.  This different low-$T$ trends indicated that for $ \phi \lesssim 0.3$,
on cooling the network becomes stretched (negative pressures), in the attempt to
preserve the connected bonded state.  This implies that
at low $T$, there is a  driving force for phase separating into a fully connected unstressed network and a gas phase.   This also suggests that the spinodal curve ends at $T=0$ around $\phi =0.3$.
At  $\phi \approx 0.3$, the packing fraction is optimal for the formation of an unstressed fully connected network at low $T$. The bond formation on cooling does not require any stretching 
and  it reverses the $T$-dependence of $P$.
(iii) Between  $0.3 \lesssim \phi \lesssim 0.38$ a minimum of $P$ appears. The existence of a minimum in $P(T)$ along isochores  evidences the presence of density anomalies (i.e. expansion on cooling along isobars) since
 points in which $(\partial P/\partial T)_V=0$, by a Maxwell
relation, coincide  with points in which $\alpha\equiv (\partial V/\partial T)_{P}=0$, i.e. with points in which density anomalies are present.
The simplicity of the model allows us to access the different contributions to $P$ and
investigate the origin of the increase of $P$ on cooling. In the PMW, apart from the trivial kinetic component contribution, the only positive component to $P$ arises from the HS interaction.
Interestingly enough, the HS component increases on cooling. Such an increase in the HS repulsion,
indirectly induced by the formation of the bonding pattern, in the range $0.30 \lesssim\phi  \lesssim 0.36$ appears to be  able to compensate the decrease in 
the bonding component of $P$. 

\begin{figure}[tbh]
\centering
%\vspace{0.10cm}
\includegraphics[width=0.45\textwidth]{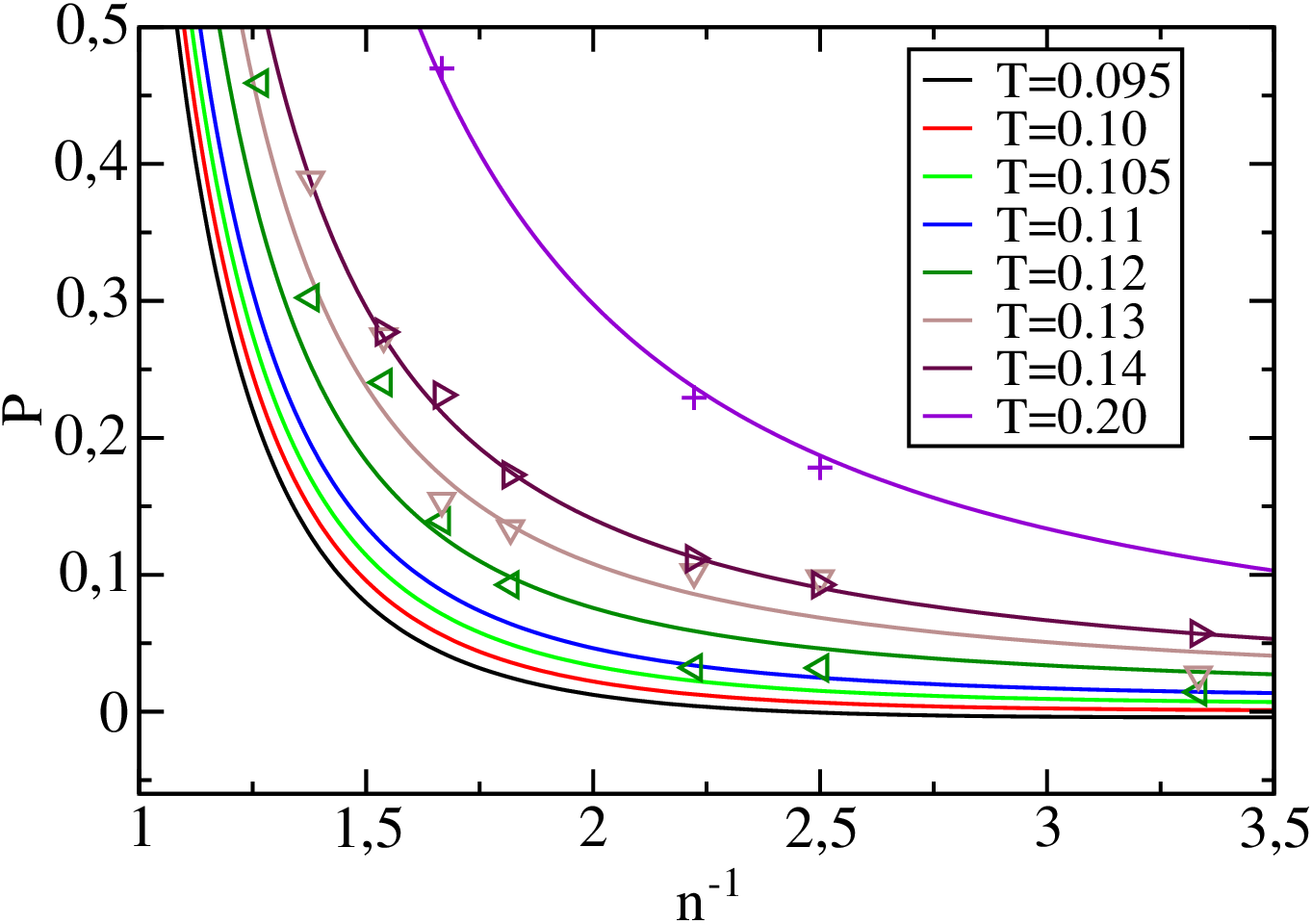}
\includegraphics[width=0.45\textwidth]{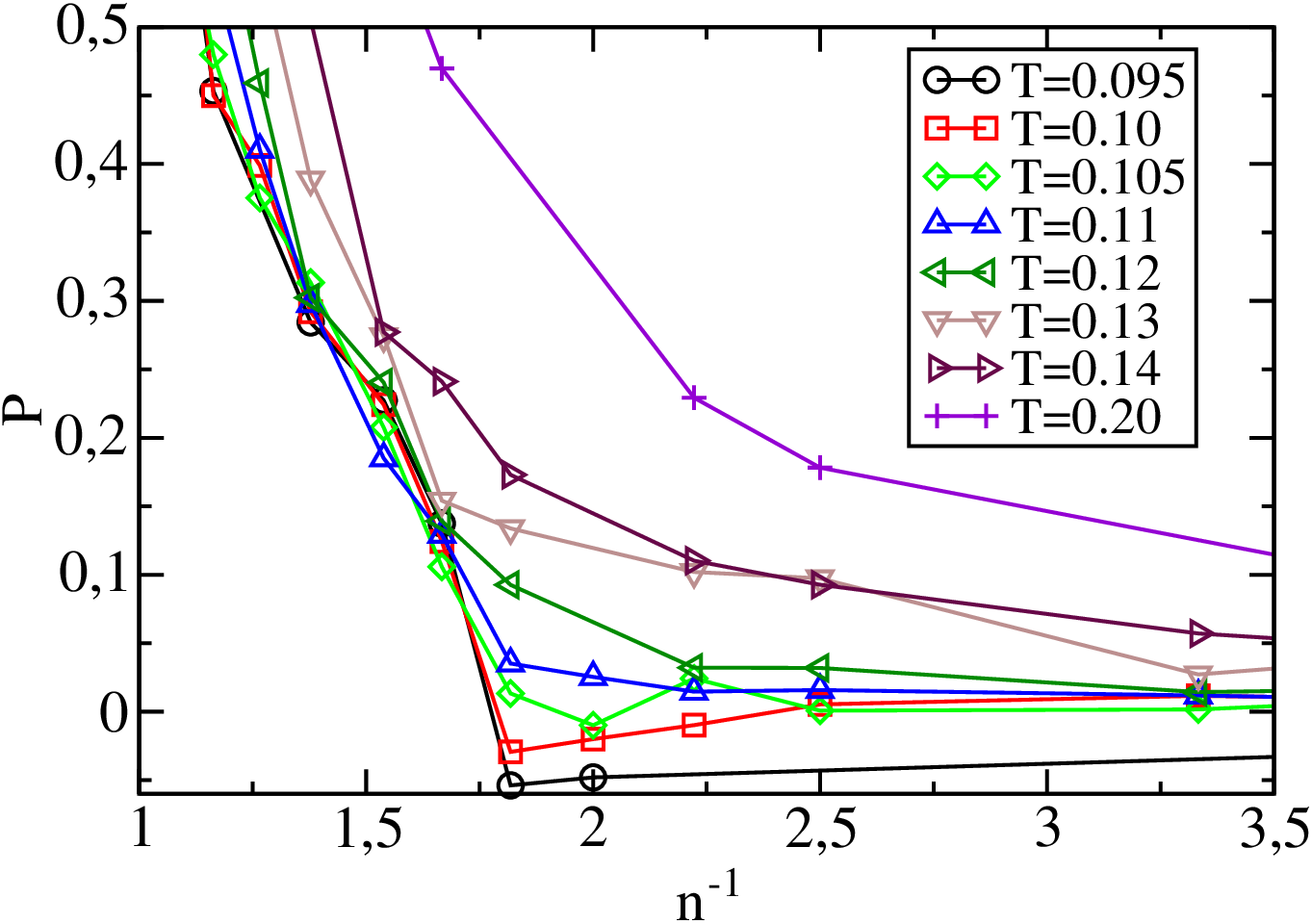}
\vspace{-0.50cm}
\caption{Isotherms of $P$ according to the Wertheim theory (top) and as calculated from the simulation data (bottom) as a function of the volume per particle $v \equiv n^{-1}$. Symbols refer to simulation data.  Note the crossing
of the different isotherms at $v=1.4$ and $1.7$. }
\label{fig:pvsrho}
\end{figure}

To confirm the presence of density anomalies it is instructive to look at the $V$ dependence of $P$ along isotherms, shown in Fig.~\ref{fig:pvsrho}. Again, the simulation data are
consistent with the Wertheim theory predictions only at large $T$ and indeed it was already noted that
no density anomalies are found within the theory\cite{Kol87a}. The simulation data also show a clear crossing of the isotherms around a volume per particle $v=1.4$ and $1.7$, corresponding to
$\phi=0.314$ and $\phi=0.38$. Again crossing is indicative of the presence of density anomalies. 
The increase of $P$ on cooling, between $\phi=0.314$ and $\phi=0.38$ suggest also a  possible emergence of a second Van der Waal-type  loop (in addition to the gas-liquid one) for $T$ lower than the one we are currently able to equilibrate. The possibility of a second critical point between two liquid phases of different densities has been discussed at length in the past\cite{Sci05a}, following the discovery of it\cite{Poo92a} in one of the first models for water\cite{Sti74a}. 

\subsection{$g(r)$} 

The PMW  radial distribution functions  for $T>0.15$,  have been reported previously\cite{Kol87a}. Here we focus on the interesting structural changes observed during the development of the bond network in $g_{OO}$ and $g_{H-LP}$, a $T$-region which was not possible to access in the previous simulations. The $g_{OO}$ provides information on the center to center particle correlation while $g_{H-LP}(r)$ contains information on the bonding and on the attractive component of the pressure.   

\begin{figure}[tbh]
\centering
%\vspace{0.10cm}
\includegraphics[width=0.45\textwidth]{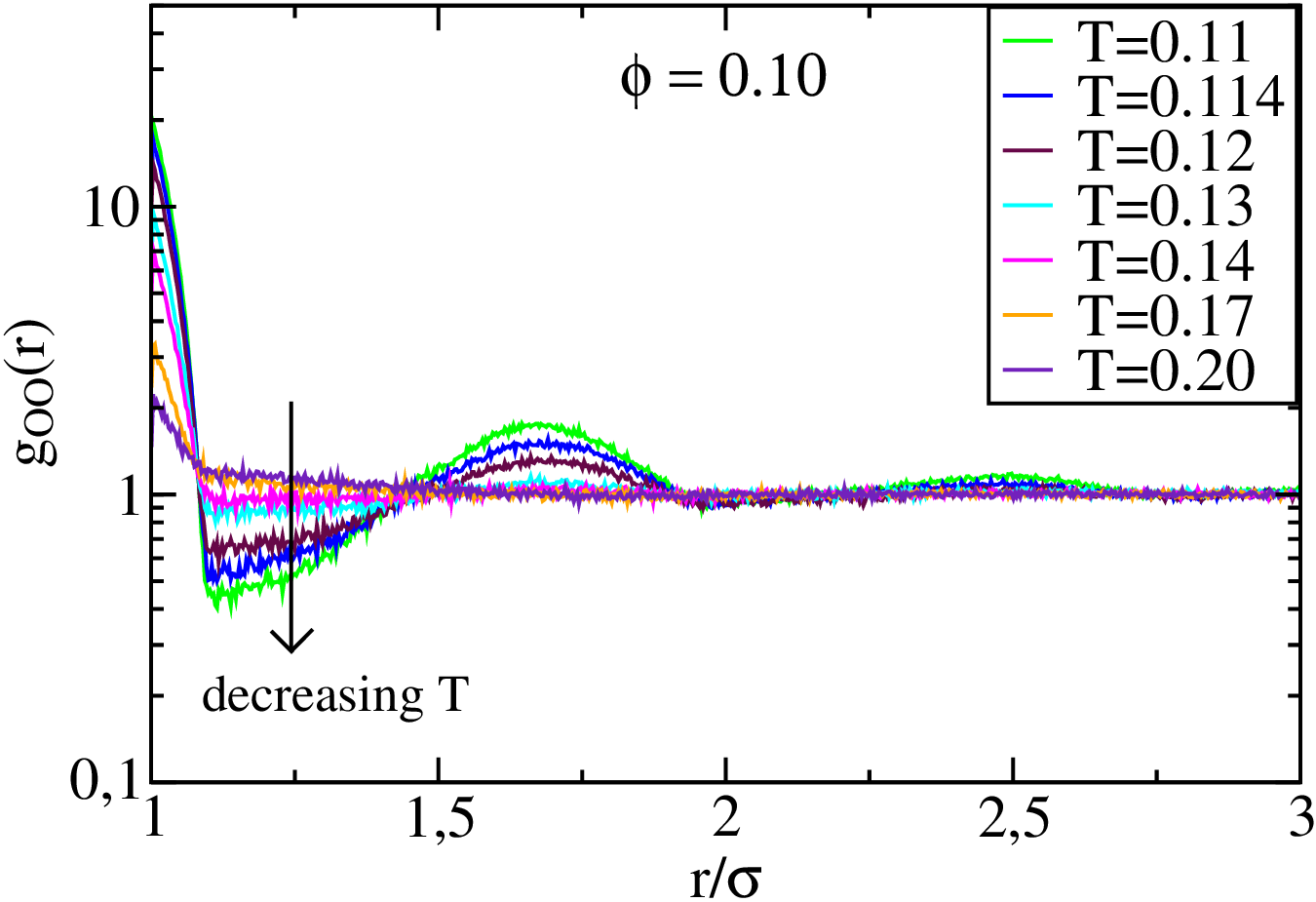}
%\vspace{-1.50cm}
\includegraphics[width=0.45\textwidth]{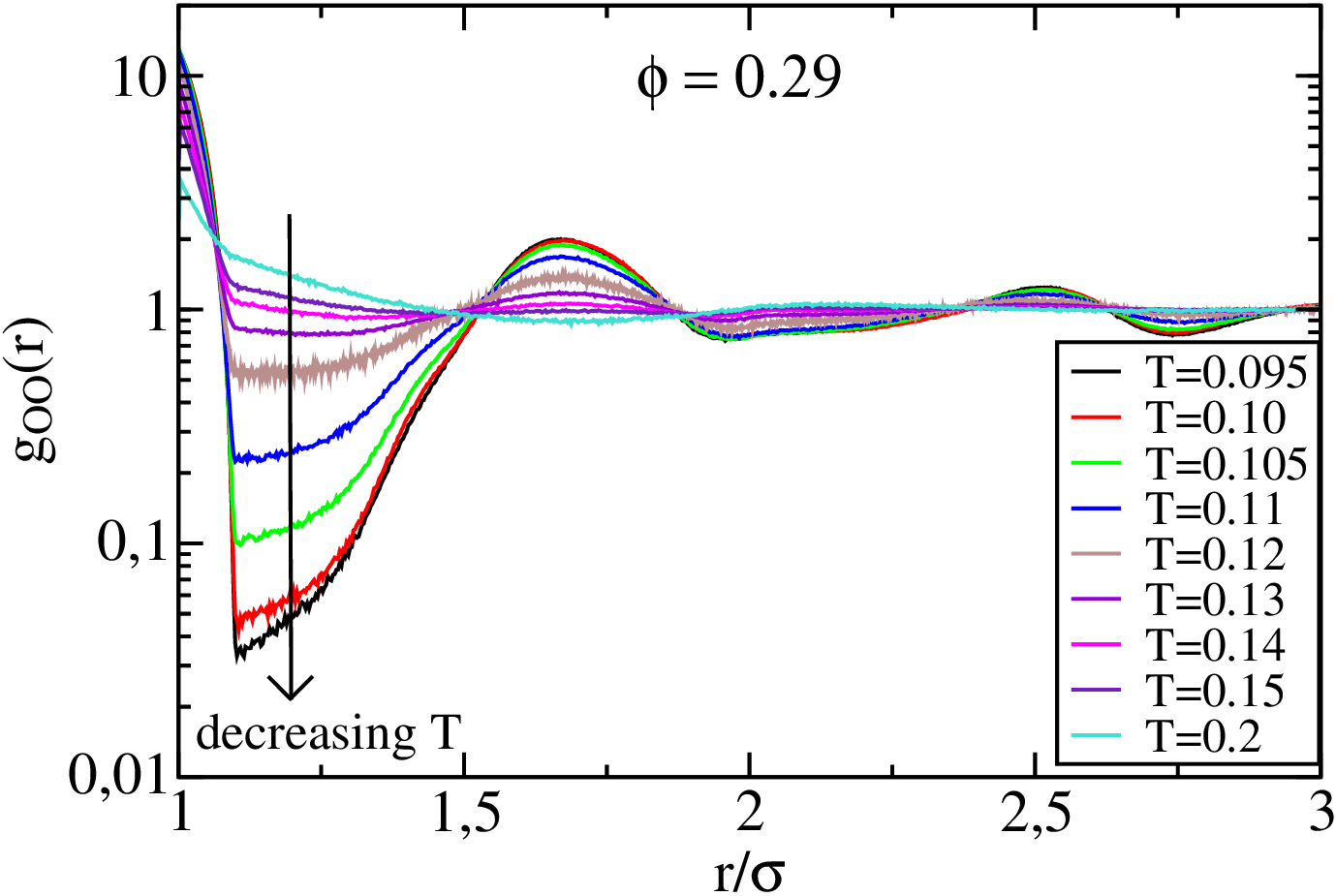}
%\vspace{-1.50cm}
\includegraphics[width=0.45\textwidth]{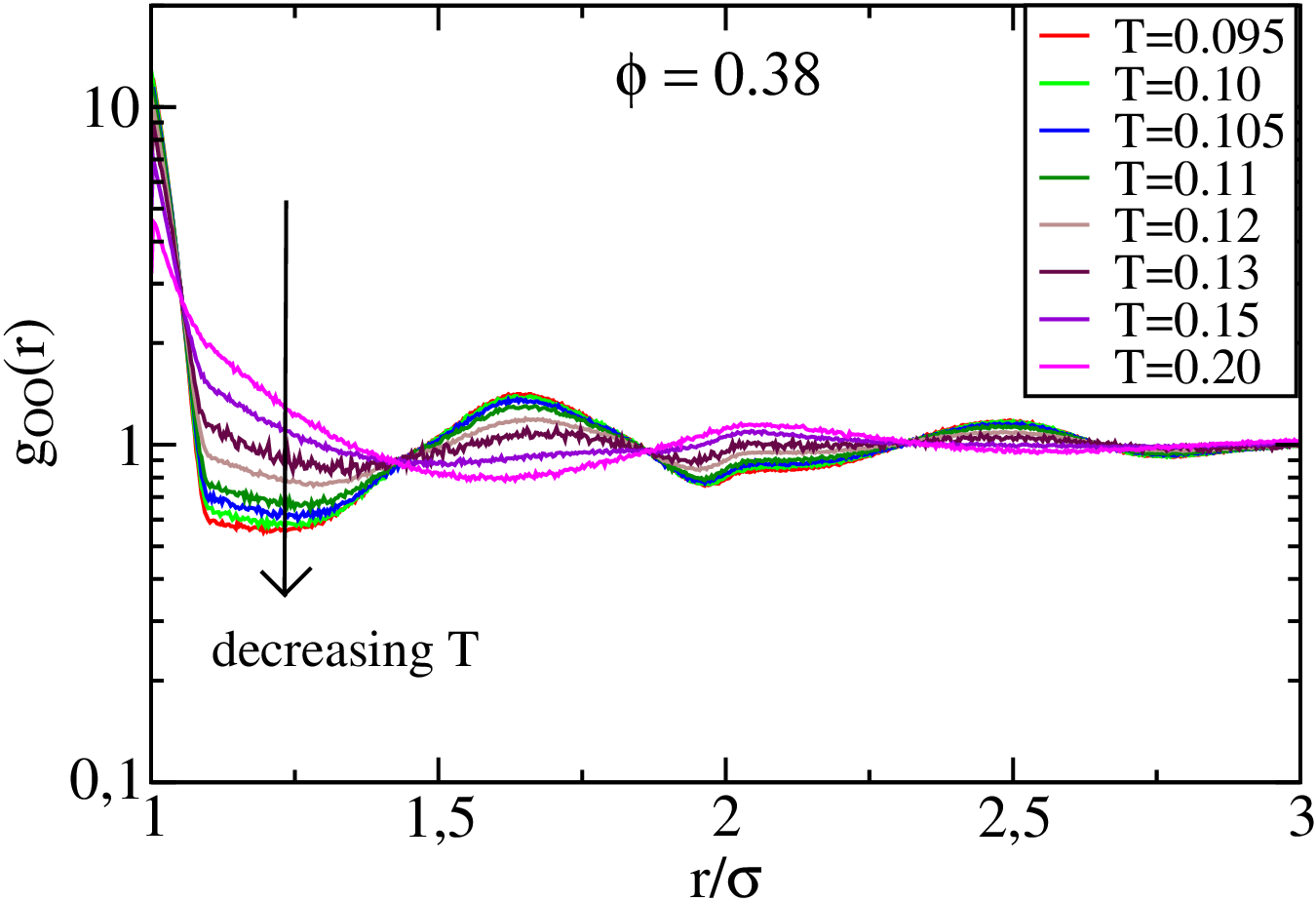}
%\vspace{0.10cm}
\caption{Particle-particle radial distribution function at
$\phi=0.105$, $\phi=0.288$, $\phi=0.380$.}
\label{fig:groo}
\end{figure}

Fig.~\ref{fig:groo} shows   $g_{OO}(r)$ at three different packing fractions. In the
interval $1<r<1.1$ the function is  highly peaked, a consequence of the distance imposed by
bonding. Outside the bonding distance ($r>1.1$),   $g_{OO}(r)$  shows significant oscillations
only at low $T$.  A peak, beside the bonding one,  is observed  at $r\approx 1.7$ corresponding to 
the characteristic distance between two particles bounded to the same central particle 
in a tetrahedral geometry. The absence of the information about the geometry of the bonding sites  in the theory of Wertheim  is responsible for the absence of the peak at
 $1.7 \sigma$ and the breakdown of the predictive ability of the Wertheim theory as soon as a particle is engaged in more than two bonds.   A few observations are in order when comparing the
 $\phi$ dependence of $g_{OO}(r)$: At low $\phi$, the tetrahedral peak at $r \approx 1.7$ 
is the only peak in  $g_{OO}(r)$. When $\phi$ approaches the optimal network density 
 a clear tetrahedral pattern develops and $g_{OO}(r=1.7)$ becomes larger than two.  The tetrahedral peaks at $\approx 1.7$ is followed by oscillations extending  up to $4\sigma$.   At even larger $\phi$,  there is still a residual signature of tetrahedral bonding at $1.7 \sigma$, but the depletion region
for $r>1.1 \sigma$ is not developed any longer, signaling a competition between the HS packing 
(which favor a peaks  at positions multiple of $\sigma$) and the local low density required by bonding.

\begin{figure}[tbh]
\centering
%\vspace{0.10cm}
\includegraphics[width=0.45\textwidth]{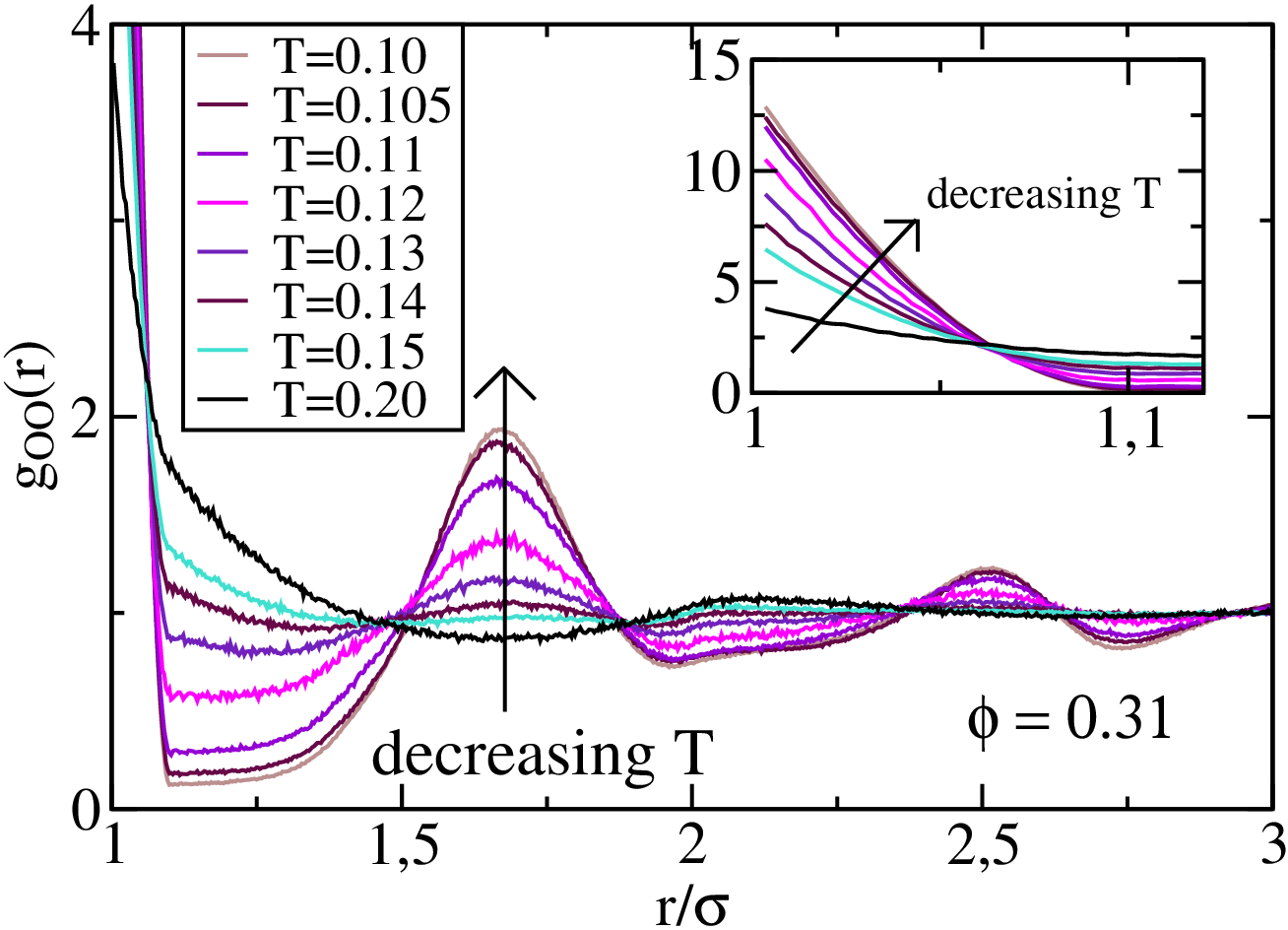}
\includegraphics[width=0.45\textwidth]{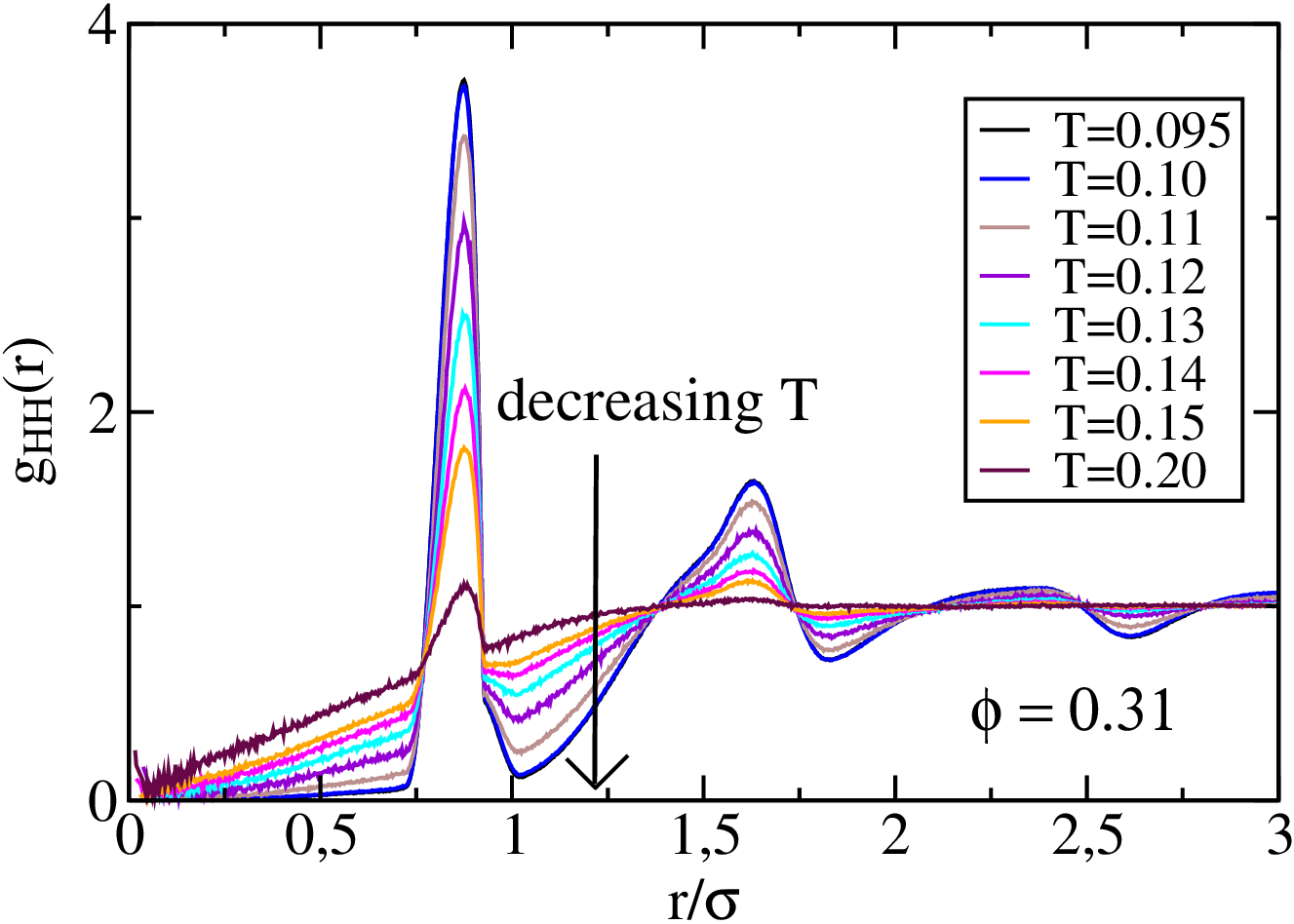}
\includegraphics[width=0.45\textwidth]{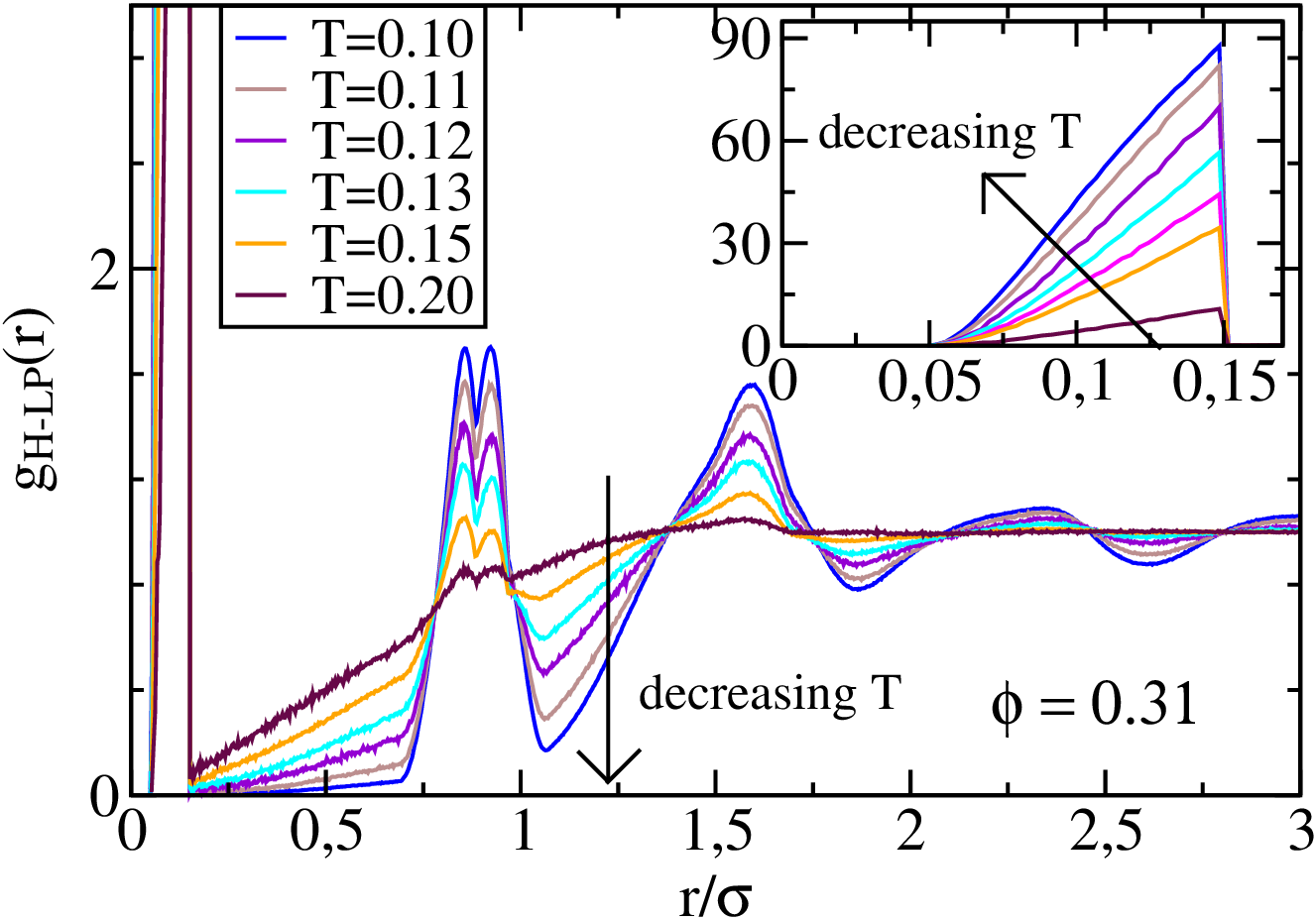}
%\vspace{0.10cm}
\caption{Radial distribution functions for $OO$, $HH$ and $H-LP$
pairs at the optimal network density $\phi=0.314$. Insets in $g_{OO}$ and $g_{_{H-LP}}$ 
provide enlargements of  the contact region. On cooling a significant structure appears, associated to the  intense bonding.}
\label{fig:grn60}
\end{figure}

Fig.~\ref{fig:grn60} compares, at $\phi=0.314$, the OO, HH and H-LP
radial distribution functions in linear scale. In all three functions, the
progressive structuring induced by the bonding is clearly evident.
Even $g_{HH}(r)$ shows very clear signs of spatial correlations,
which are induced by the tetrahedral geometry of the bonding
and by the geometry by which the bonding  between $H$ and $LP$ propagates. Indeed,
in the PMW model the interaction between different $H$ sites is zero.  

\subsection{$S(q)$}  
The structure factor of the system, defined in term of the particle's 
center coordinates $\vec r_i$ as,
\begin{equation}
S(\vec q)=<\frac{1}{N}\sum_{i=1}^N e^{i \vec q \cdot (\vec r_i - \vec r_j)}>
\end{equation}
provides information on the wave vector dependence of the density fluctuations. In isotropic systems, $S(q)$ is function of the modulus $q$.
 The behavior of $S(q)$ at small $q$ provides indication on the
phase behavior, since an increase of $S(q)$ at small $q$ indicates the development of inhomogeneities with length-scale comparable to the system size studied. As an indicator of the location of the
phase boundaries (of the liquid-gas spinodal line), we estimate
the locus of points in $(T,\phi)$ where $S(q)$ for the particles centers becomes larger than 5 at small $q$. This locus is reported in Fig.~\ref{fig:phase}. For $\phi \gtrsim  0.28$ $S(q)$ does not show any sign of growth at small $q$ in the region of $T$ where equilibration is feasible, being characterized by values of $S(q)$ at small $q$ of the order of $0.1$.  This confirms that, at this packing fraction, there is no driving force for phase separation, since the average density has reached a value such that  the formation of a fully connected network of bonds does not require a local increase of the packing fraction.  It is also important to stress that at  $\phi =0.288$,  at the lowest studied $T$, the average number of bond per particle is $3.8$, and hence the system is rather close to its ground state and no more significant structural changes are expected on further cooling.

\begin{figure}[tbh]
\centering
%\vspace{0.10cm}
\includegraphics[width=0.45\textwidth]{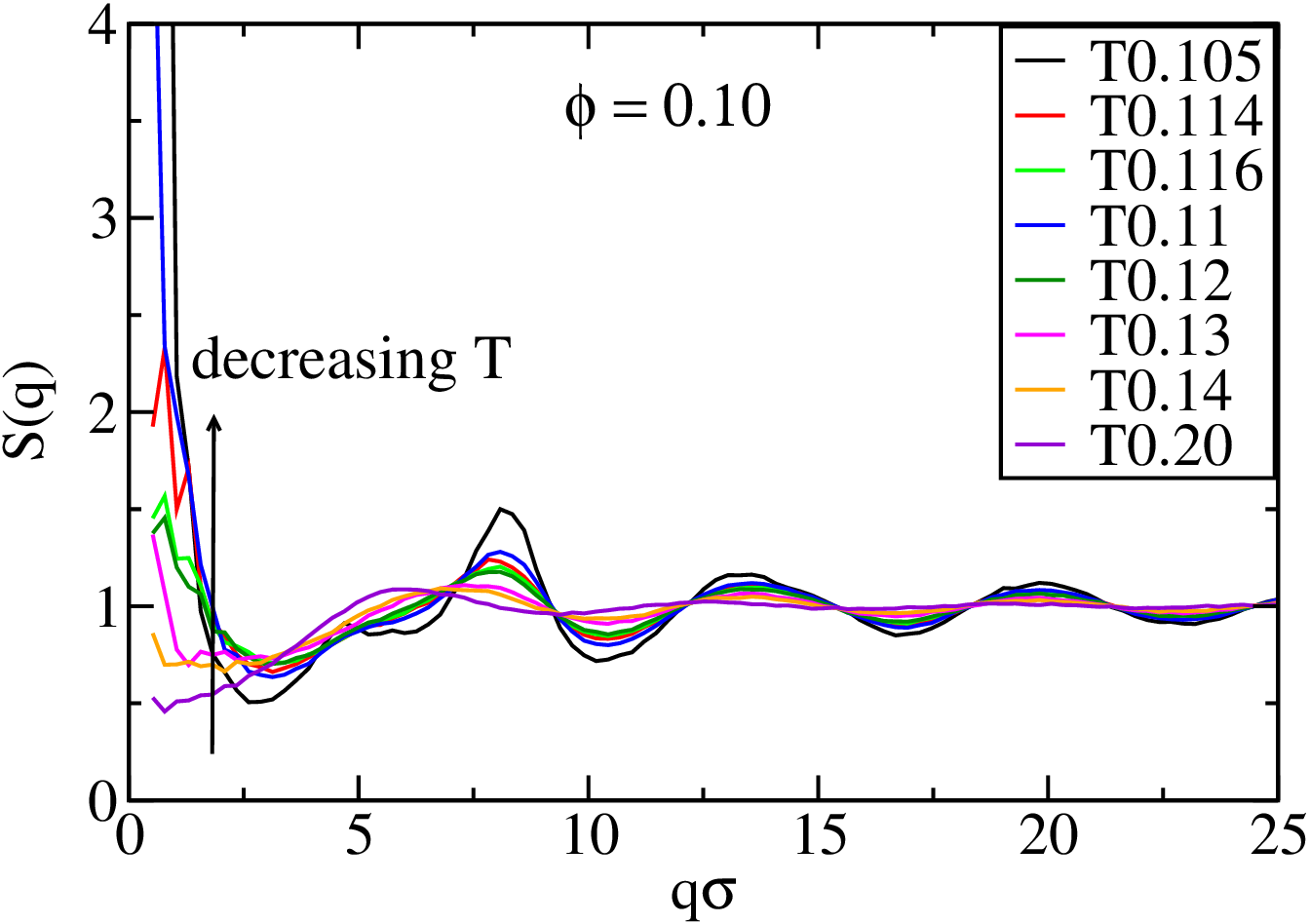}
\vskip 0.8cm
\includegraphics[width=0.45\textwidth]{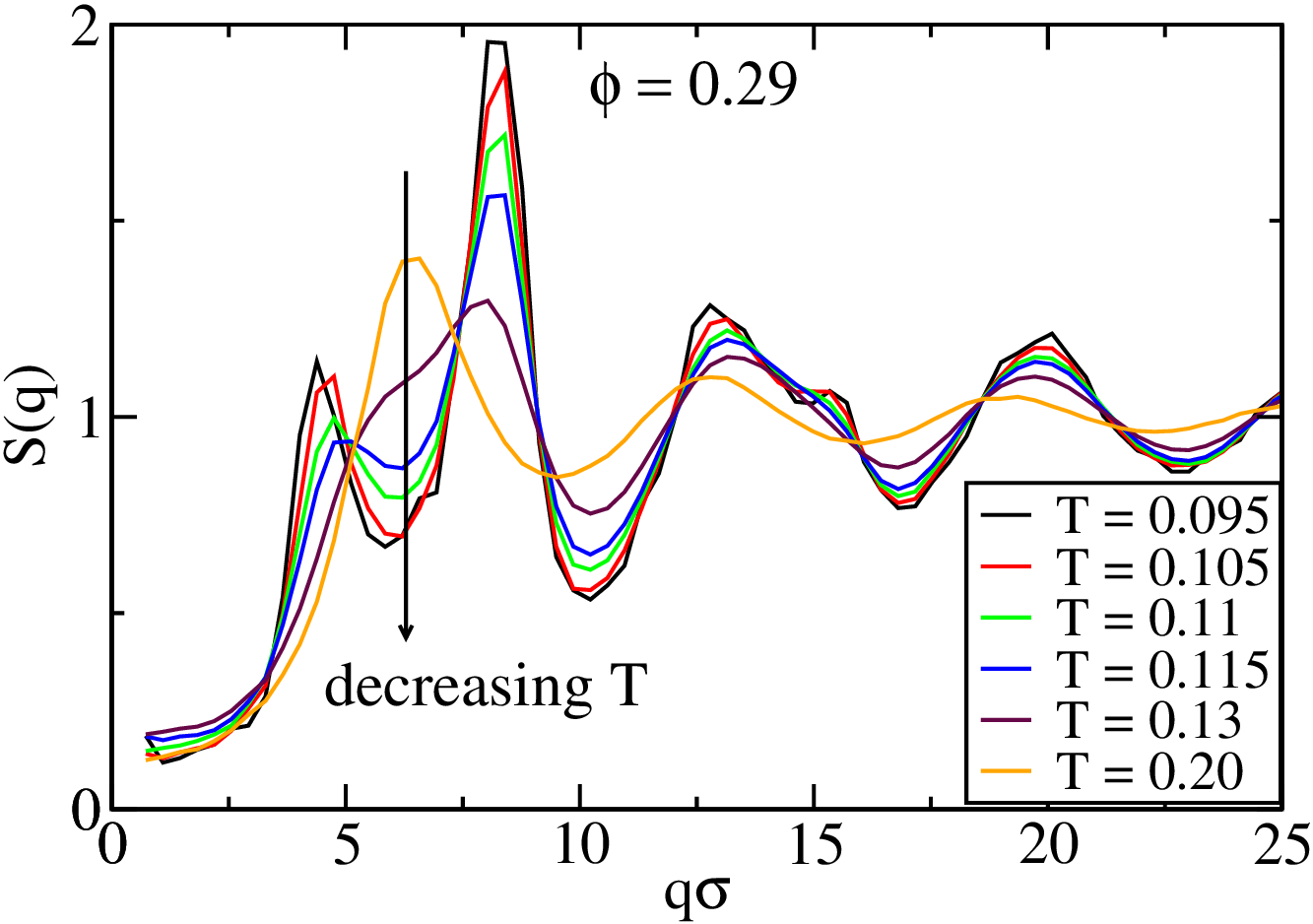}
\vskip 0.25cm
\includegraphics[width=0.45\textwidth]{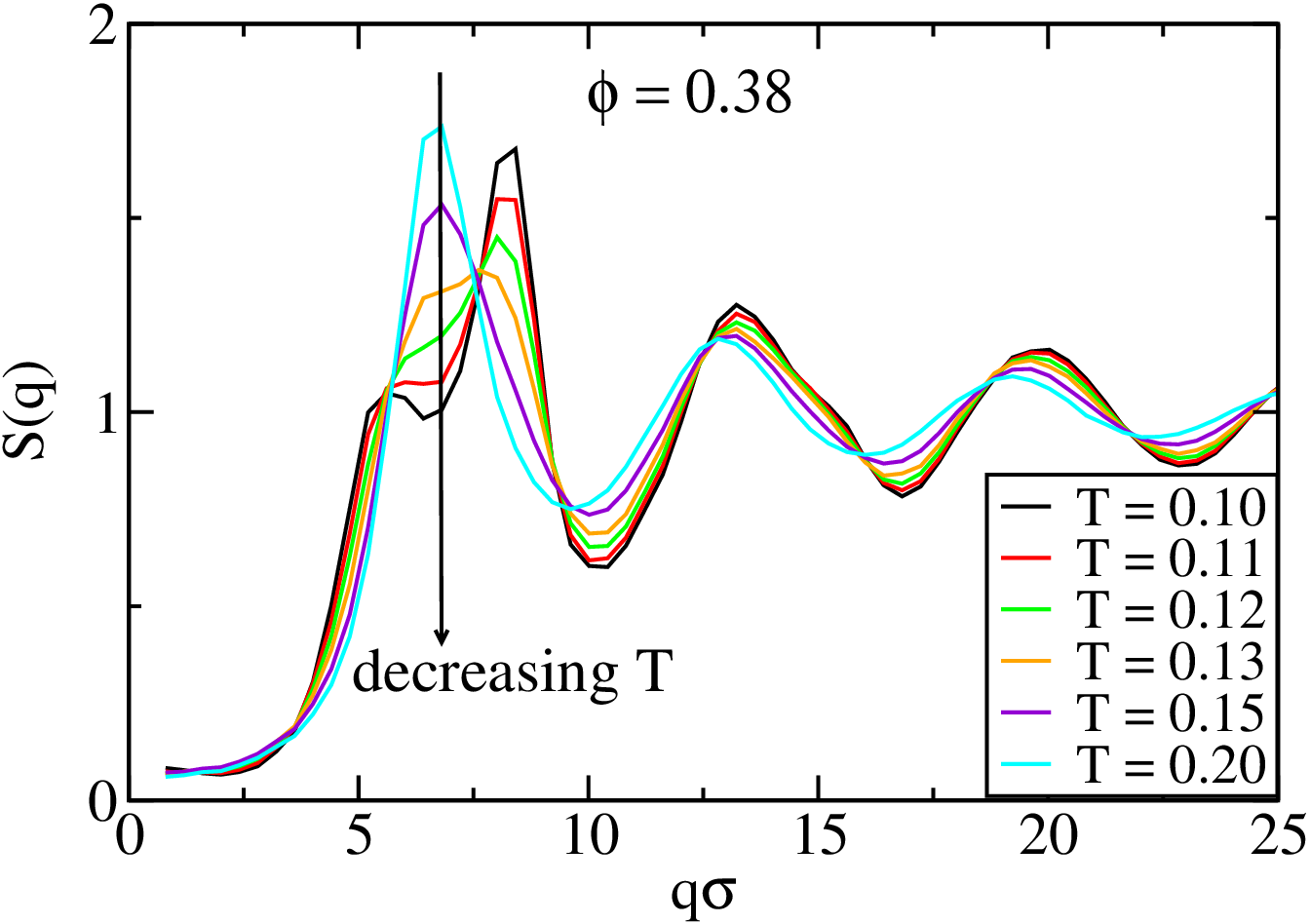}
%\vspace{0.10cm}
\caption{Particle-particle structure factor at $\phi=0.105$, $\phi=0.288$, $\phi=0.3858$. 
Note that at $\phi=0.105$, an intense signal develops at small $q$, related to the approach to the spinodal instability. Small $q$ intensity is completely missing at the
higher $\phi$ shown. }
\label{fig:sqoo}
\end{figure}

Fig.~\ref{fig:sqoo} shows $S(q)$ at $\phi =0.105$, $\phi =0.288$ and $\phi=0.385$. 
The $\phi =0.105$ case has been chosen to show the
significant increase in $S(q)$ associated to the approach of the
spinodal curve.  The case $\phi=0.288$ shows both the absence of a small $q$-vector divergence and the clear development of
the typical $q$-pattern of tetrahedral networks. On cooling the
peak at $q\sigma=2\pi$, characteristic of excluded volume interactions splits in two parts. A pre-peak around $q\sigma \approx 5$ and an intense peak  around $q\sigma \approx 8$.  The case $\phi=0.385$ confirms that
the packing fraction is now so high that a full tetrahedral network cannot develop and the splitting of the main peak in two distinct components
is very weak and visible only at the slowest investigated $T$.

\subsection{Percolation}

The PMW, as all other models based on HS and SW interactions,
is particularly suited for calculation of bond properties, since a
bond between particle $i$ and $j$ can be unambiguously defined when
the pair interaction energy between $i$ and $j$ is $-u_0$. In the case of continuous potentials such a clear cut bond definition is not possible and several alternative propositions have been put forward\cite{hill,Con04aJPCM}.
We focus here on the connectivity properties of the equilibrium
configurations.  We use standard algorithms to partition particles into clusters.   Configurations are considered percolating when,
accounting for periodic boundary conditions, an infinite cluster is
present.  More explicitly, to test for percolation,
the simulation box is duplicated in all directions and the ability of the largest cluster to span the replicated system is controlled. If the cluster in the simulation box does not connect with its copy in the duplicated system then the configuration is assumed to be non-percolating.
The boundary between a percolating and a non-percolating
state point has been defined by the probability of observing infinite
clusters in 50$\%$ of the configurations. The resulting percolation line
is reported in Fig.~\ref{fig:phase}. State points on the right side of the
line are characterized by the presence of an infinite cluster. Still,  
at this level of definition, percolation is a geometric measure and it does not provide any information on the lifetime of the percolating cluster.

The percolation line, like in simple SW potentials, crosses the spinodal curve very close to the critical point.  Differently from the SW, the percolation locus  does not extend to infinite $T$, since at high $T$, even at  large $\phi$, the reduce particle surface available for bonding 
prevents the possibility of forming a spanning network with a random 
distribution of particles orientations.  Along the percolation line, about 1.5 bonds per particle are observed, with a small trend towards an increase of this number on decreasing $\phi$.  In terms of
bond probability $p_b$, this correspond to $p_b \approx 0.375$, not too different from the
bond percolation value of the diamond lattice, known to be $0.388$\cite{Sta92book}.

\section{Dynamics}

Thermodynamic and static properties of the PMW presented in the previous section clarify the location of the regions in which the bond network forms, the
region where the liquid-gas phase separation takes place and the region at  high $\phi$ where packing phenomena start to be dominant.  In the following we present a study of  the diffusion properties
of the model in the phase diagram, with the aim of locating the limit of stability of the liquid state imposed by kinetic (as opposed to thermodynamic) constraints.

\subsection{$MSD$}

\begin{figure}[tbh]
\centering
\vspace{0.10cm}
\includegraphics[width=0.45\textwidth]{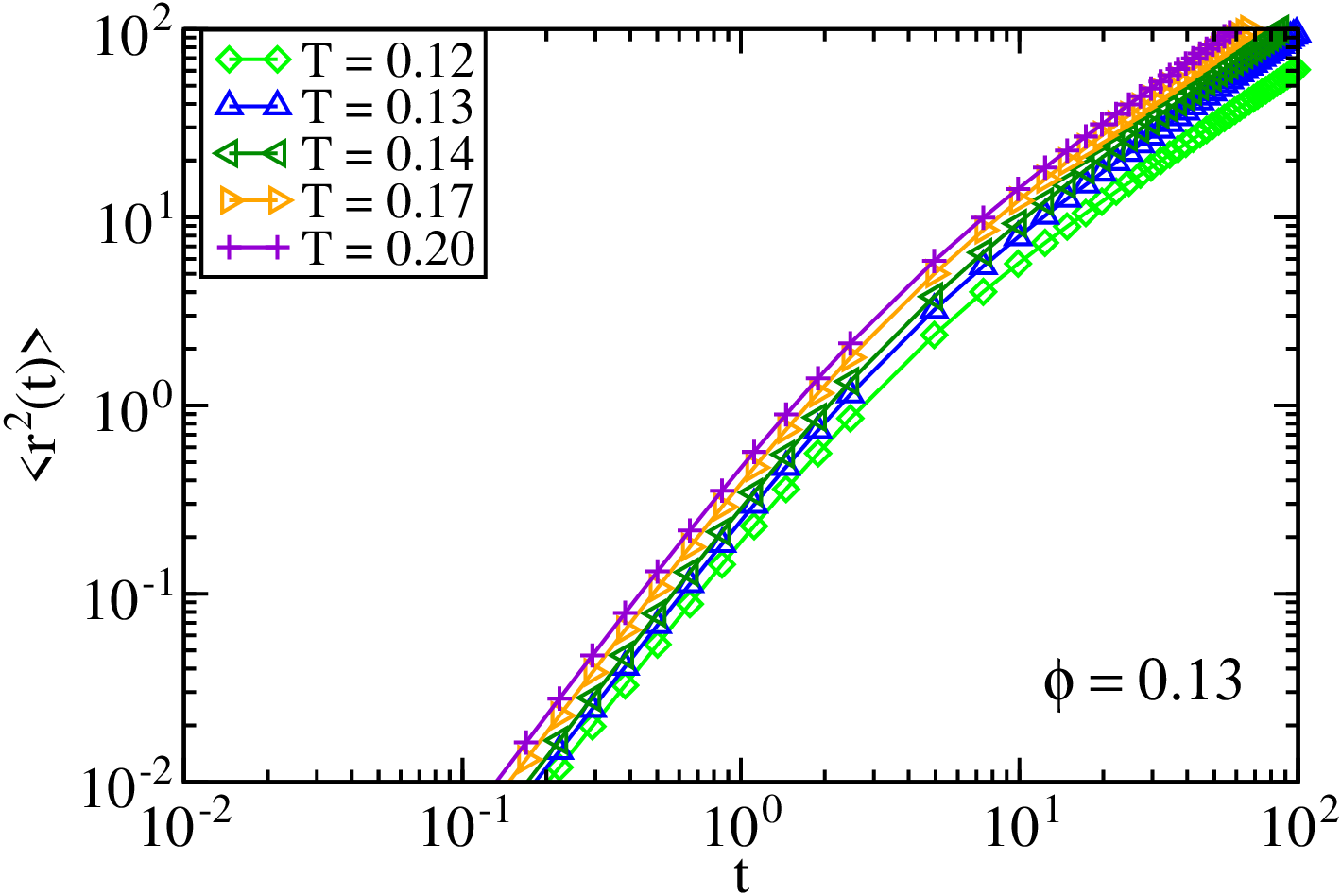}
\includegraphics[width=0.45\textwidth]{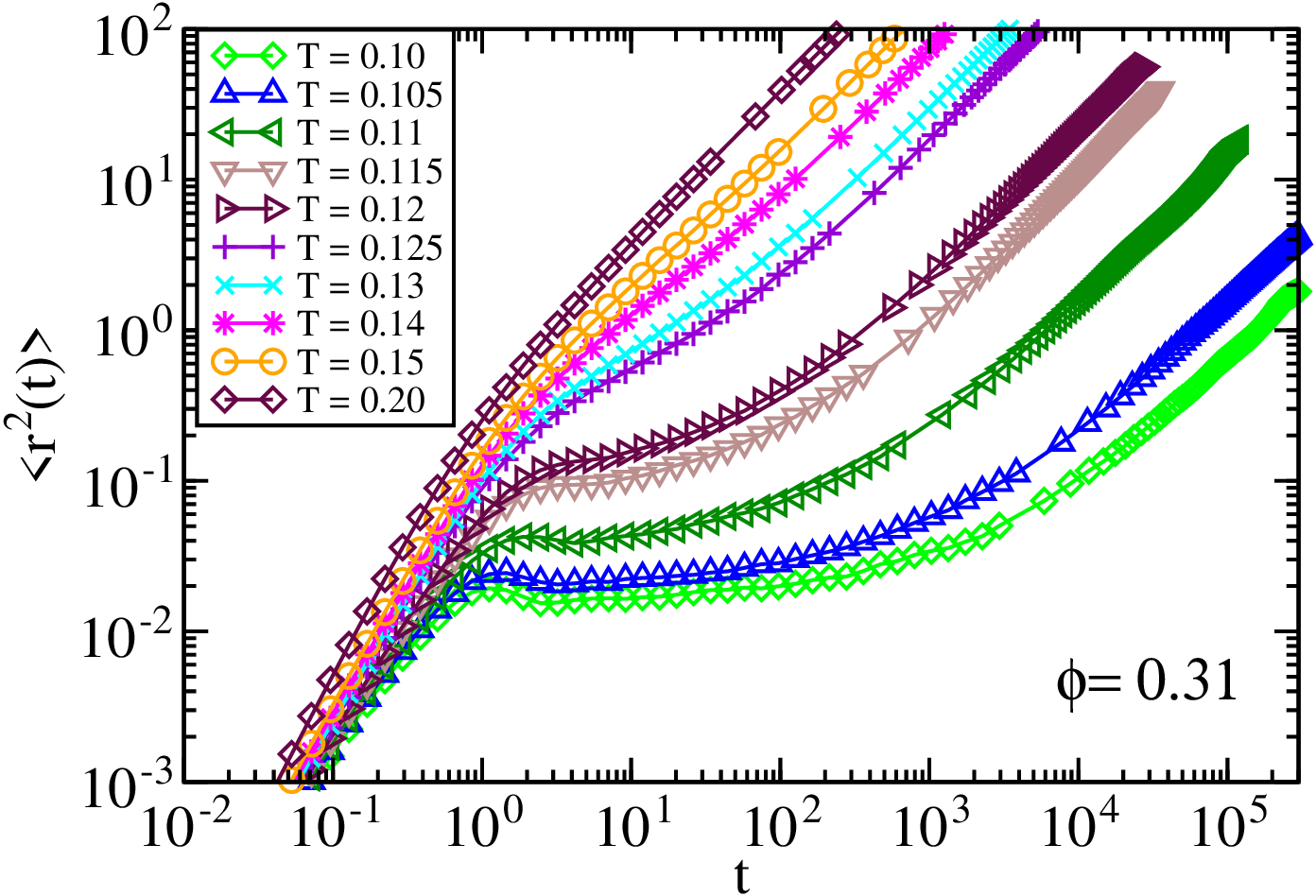}
\includegraphics[width=0.45\textwidth]{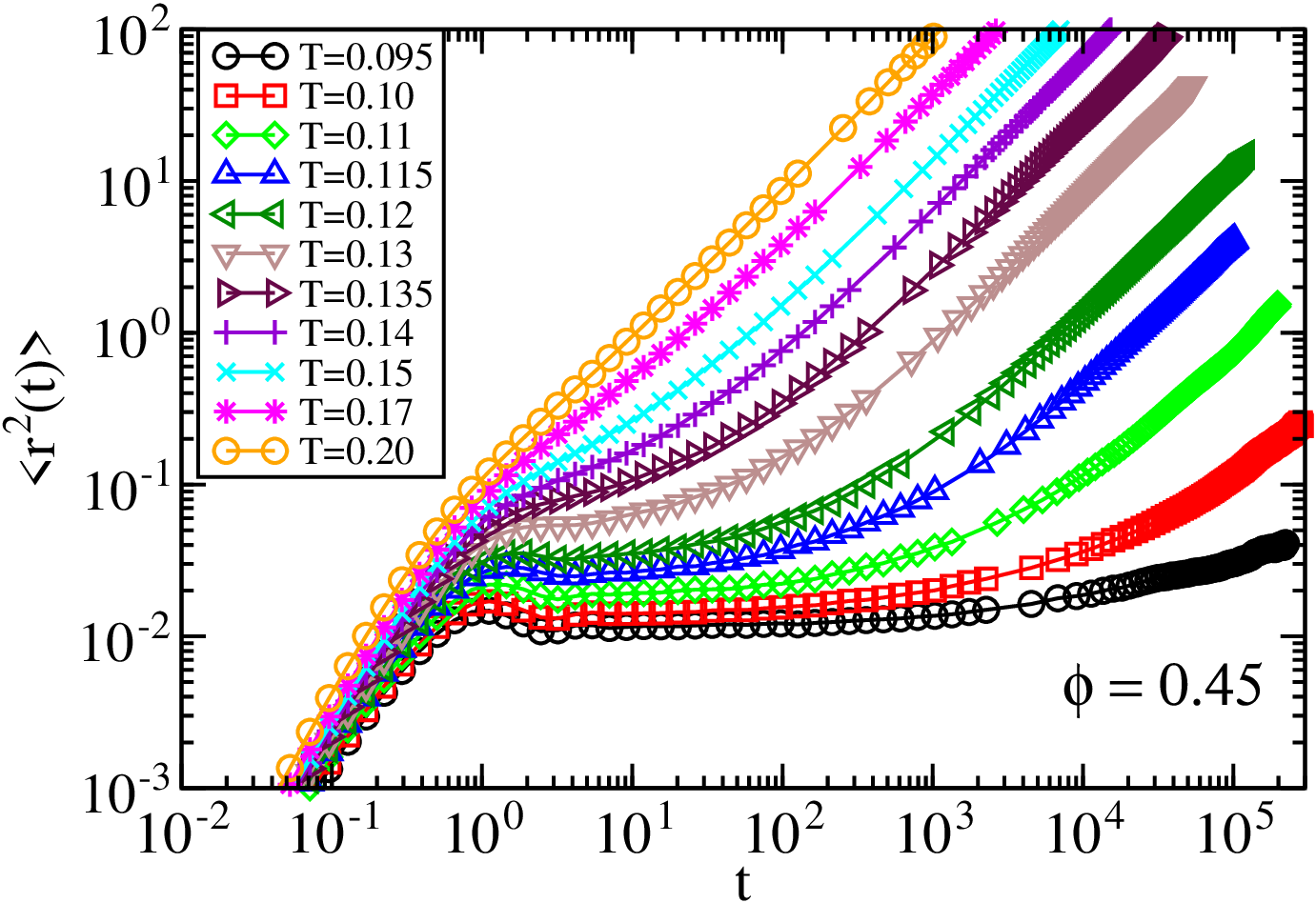}
%\vspace{0.10cm}
\caption{Mean square displacement for different $T$s, at three different $\phi$ values. 
Top: $\phi=0.131$, Center: $\phi=0.314$, Bottom: $\phi=0.450$.}
\label{fig:msd}
\end{figure}

We focus on the mean square displacement $<r^2(t)>$ of the particle centers, as a function of $T$ and $\phi$, calculated from the newtonian dynamic trajectories. Fig.~\ref{fig:msd} shows $<r^2(t)>$ for a few selected isochores. For short time $<r^2(t)> = <v_T^2> t^2$, where  $<v_T^2>=3/2 k_BT$ is the thermal velocity. At high $T$, the short-time ballistic behavior crosses over to a
diffusion process ($<r^2> \sim t$) directly. At low $T$, the ballistic short-time and the diffusive long time 
laws are separated by an intermediate time window in which $<r^2(t)>$  is approximatively constant, an indication of particle caging.

\begin{figure}[tbh]
\centering
\vspace{0.10cm}
\includegraphics[width=0.45\textwidth]{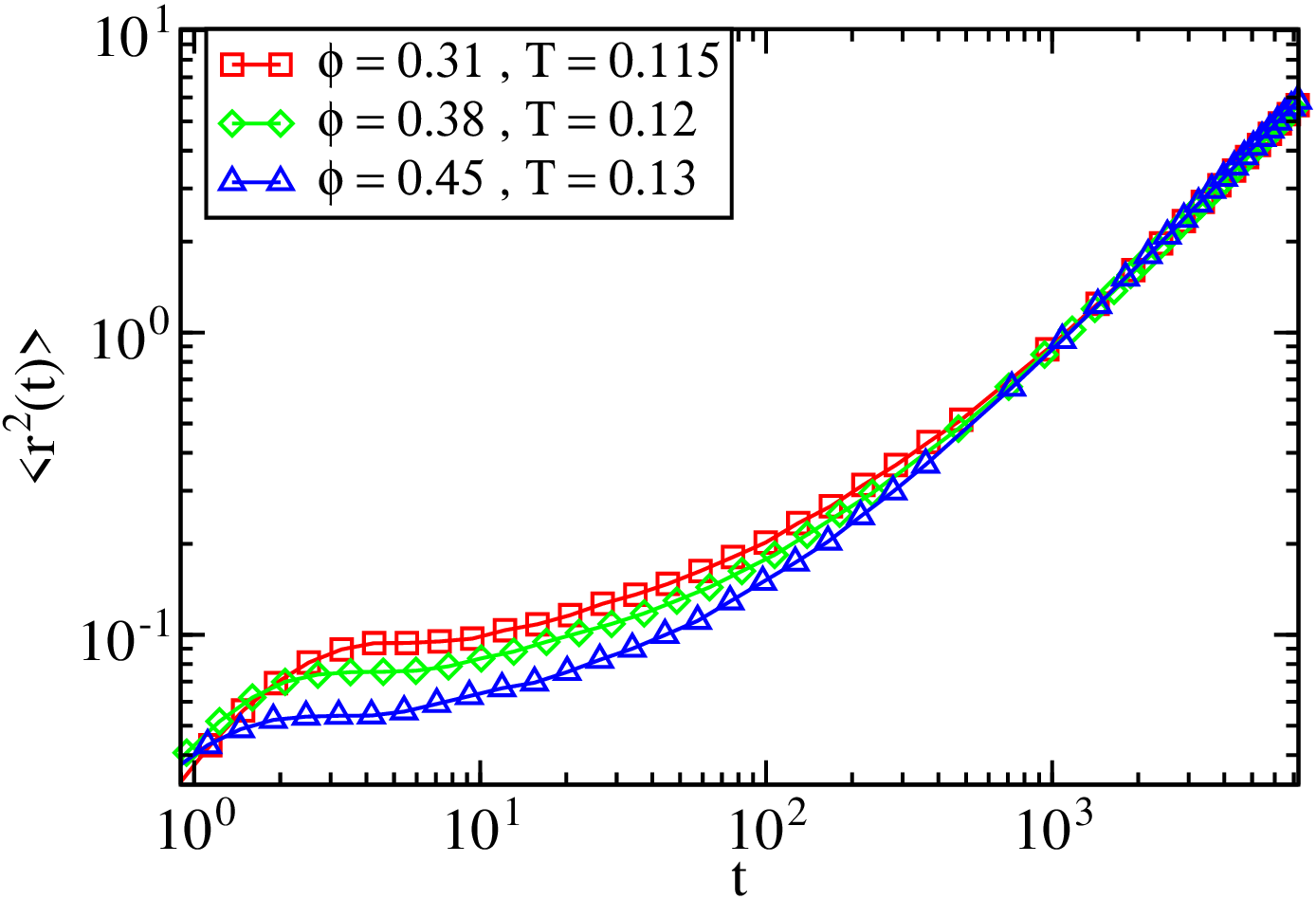}
%\vspace{0.10cm}
\caption{Mean square displacement along a constant $D$ path. 
Note the $\phi$ dependence of the plateau at intermediate times, which provides an
estimate of the caging length. }
\label{fig:isoD}
\end{figure}

Several features of  $<r^2(t)>$ are worth pointing:
(i) For $\phi \lesssim 0.209$, the spinodals are encountered on cooling 
before the caging process is visible.   The phase separation process sets in well before
particles start to feel the caging process.  
(ii) The static percolation curve reported in Fig.~\ref{fig:phase} has no
effect on dynamics. There is no  dynamic arrest
at the static percolation transition.
(iii) For $\phi$ such that a well developed tetrahedral network can form,
it is possible to cool the system down to temperatures at which, on the scale of simulation, arrest is observed, in the absence of any phase separation.  $<r^2(t)>$ develops a clear intermediate
region where only the dynamic inside the cage is left. At this $\phi$, the caging  is not associated 
to excluded volume interactions, but to the formation of energetic bonds\cite{Sci96b}.
(iv) The plateau value in $<r^2(t)>$  is a measure of the localization
length induced by the cage.  To visualize the $\phi$ dependence of
the localization length, we show in Fig.~\ref{fig:isoD}  $<r^2(t)>$
for three different state points ($\phi$-$T$) with the same  long time diffusivity.  The cage length is
always significantly larger than the typical $HS$ value ($<r^2(t)> \sim 0.01)$ and grows  on decreasing $\phi$. 

\subsection{Diffusion Coefficient}

\begin{figure}[tbh]
\centering
\vspace{0.10cm}
\includegraphics[width=0.45\textwidth]{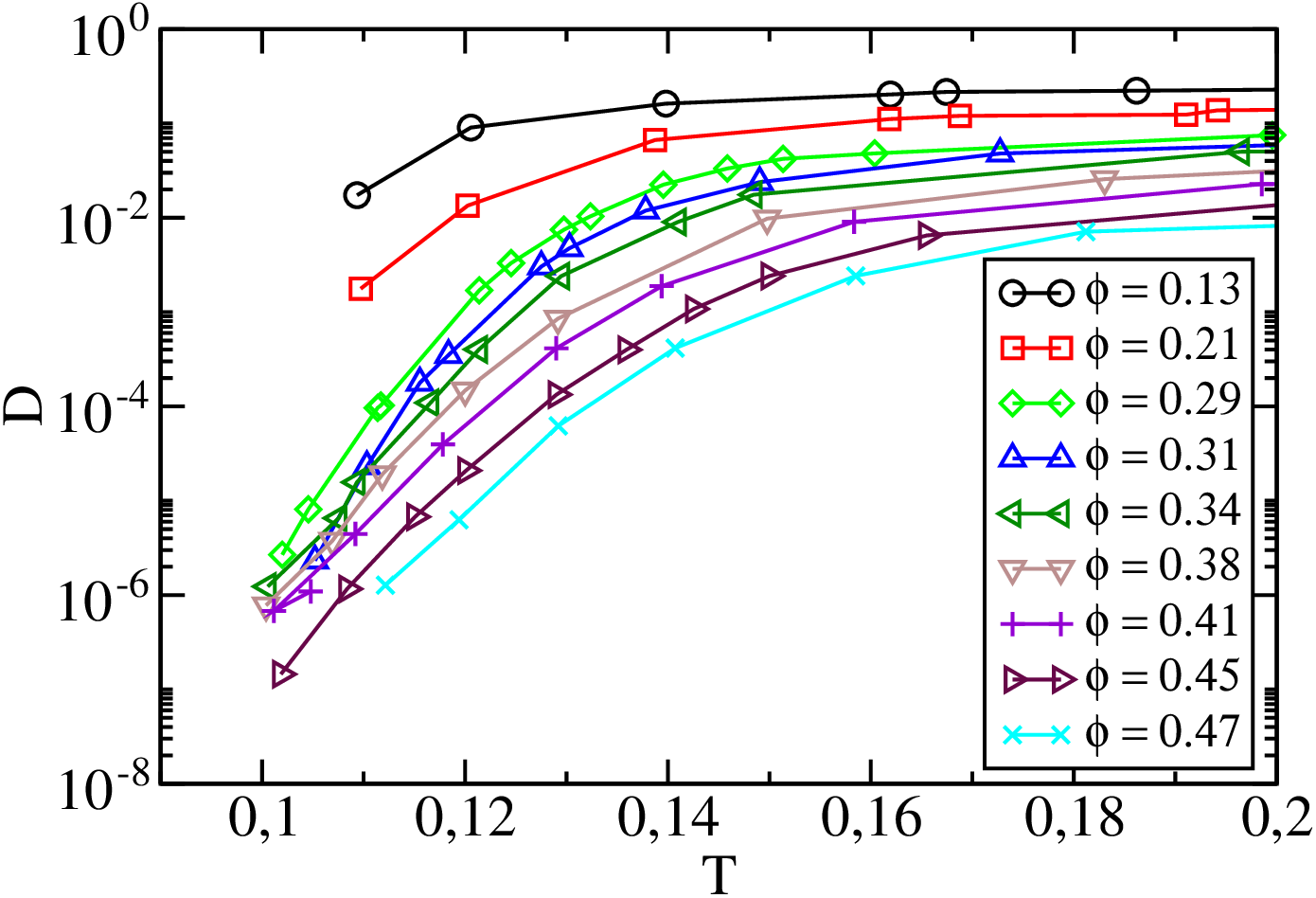}
\includegraphics[width=0.45\textwidth]{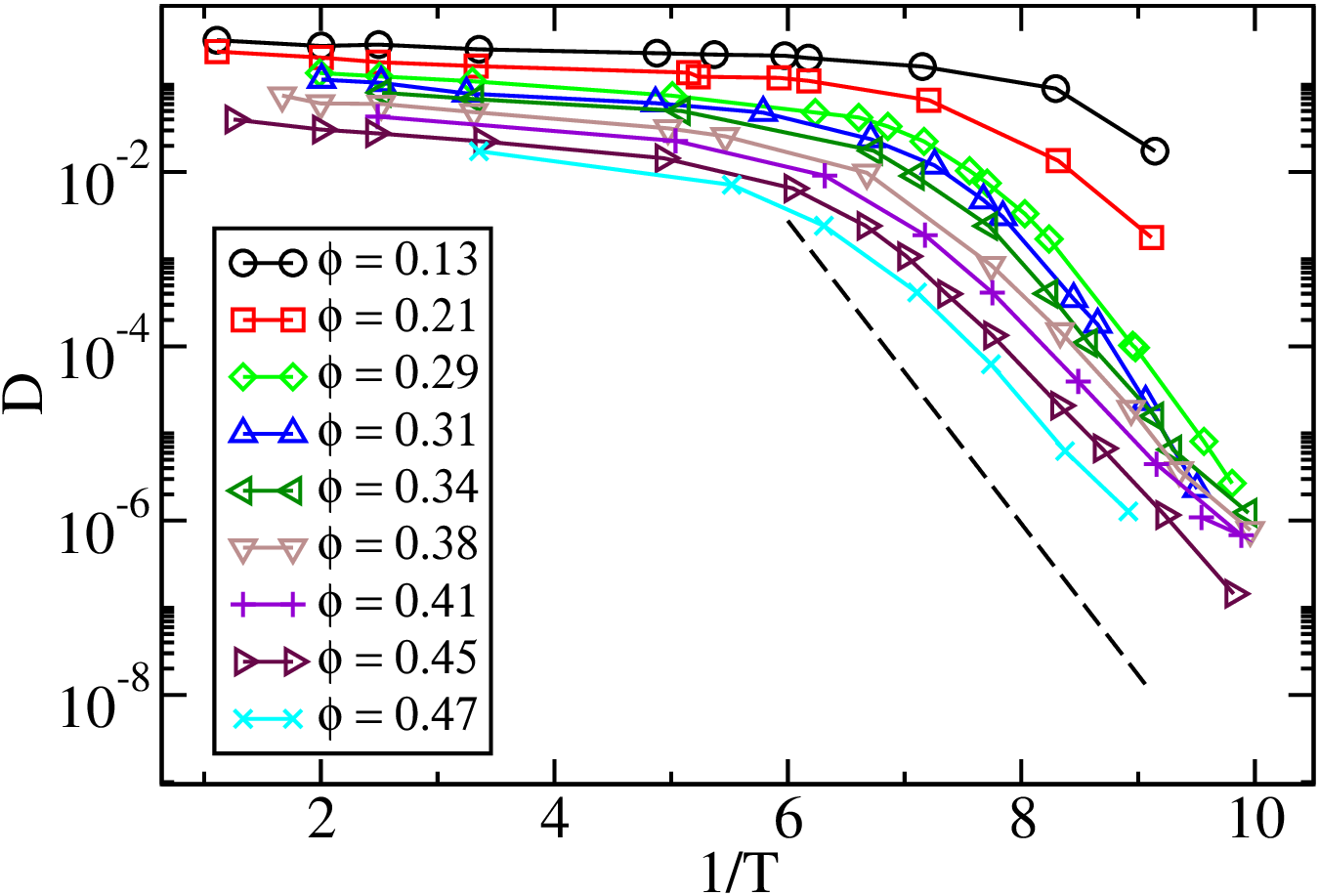}
%\vspace{0.10cm}
\caption{Temperature dependence of the diffusion  coefficient along isochores. The dashed line is an 
Arrhenius dependence with activation energy equal to $4$.}
\label{fig:DMD}
\end{figure}

The long time limit of $<r^2(t)>$ is, by definition,  $6 D t$, where $D$ is the
diffusion coefficient. The $\phi$ and $T$ dependence of $D$ is shown in
Fig.~\ref{fig:DMD}.  We show $\log(D)$ both vs. $T$ and vs $1/T$.
Again, a few considerations are in order:
(i) The range of $D$ data covers about five orders of magnitude. The data for $\phi<0.24$
are limited in $T$ by the phase separation process, while the data for $\phi>0.26$ are
limited by computational resources, since equilibration can not be reached within several months
of calculations.
(ii) Data for   $\phi>0.26$  crosses around $T\approx 0.105$, suggesting a non monotonic behavior of the $\phi$ dependence of the dynamics. 
(ii) The early decay of $D$ with $T$ can be  described
with a power-law  $|T-T_{MCT}|^{\gamma}$.  Power law fits, limited
to the region of $T$ between $T=0.11$ and $T=0.15$, cover the first
two-three orders of magnitude in  $D$, in agreement with previous studies of more detailed
models for water\cite{Gal96b,Sci96b,Sta99a} and with the previously proposed MCT interpretation
of them\cite{Gal96b,Fab99b,Sci97b,Fab98a}.
(iii) A cross-over to an Arrhenius activated dynamics is observed at low $T$. Activated processes become dominant in controlling the slowing down of the dynamics. The activation energy  is $\approx 4$ close to the
optimal network $\phi$, suggesting that at low $T$ diffusion requires breaking of
four bonds. The cross-over from an apparent power-law dependence to an Arrhenius dependence 
has also been observed in simulations of other network forming liquids, including 
silica\cite{Hor99a,Voi01a} and more recently water\cite{Xul05aPNAS}.  
The low $T$ Arrhenius dependence also suggests that in the region where bonding is responsible for caging  the
vanishing $D$ locus coincides with the $T=0$ line.

Particularly interesting is the behavior of $D(\phi)$ along isotherms.
An almost linear dependence at small $\phi$ (up to $\phi=0.235$)
is followed by a non monotonic behavior. Below $T=0.11$,
a diffusion anomaly is observed in the $T$ and $\phi$ region where
the tetrahedral network develops. Around $\phi=0.34$ an isothermal compression of the system generates a speed up of the dynamics. Above $\phi \approx 0.35$, 
$D$ starts to decrease again on increasing packing.  
Diffusivity anomalies of the type observed in the PMW are found in
several tetrahedral network forming liquids, including water\cite{Sca00a}. The explanation for this counterintuitive $\phi$ dependence of the dynamics is to be found in the geometric constraints requested by the tetrahedral bonding requiring an open local structure. Increasing $\phi$ destroys the local bonding order with a resulting speed up of the dynamics. 

%The $T_{MCT}(\phi)$ line, resulting from the power-law fits along isochores, is shown in Fig.\ref{fig:phase}. Along isotherms, reliable power-law fit are more difficult, since at large $\phi$ crystallization intervenes. This is not unexpected, since at large $\phi$ values the bonding pattern becomes less relevant and the HS crystal become favored at all but low $T$.

\begin{figure}[tbh]
\centering
\vspace{0.10cm}
\includegraphics[width=0.45\textwidth]{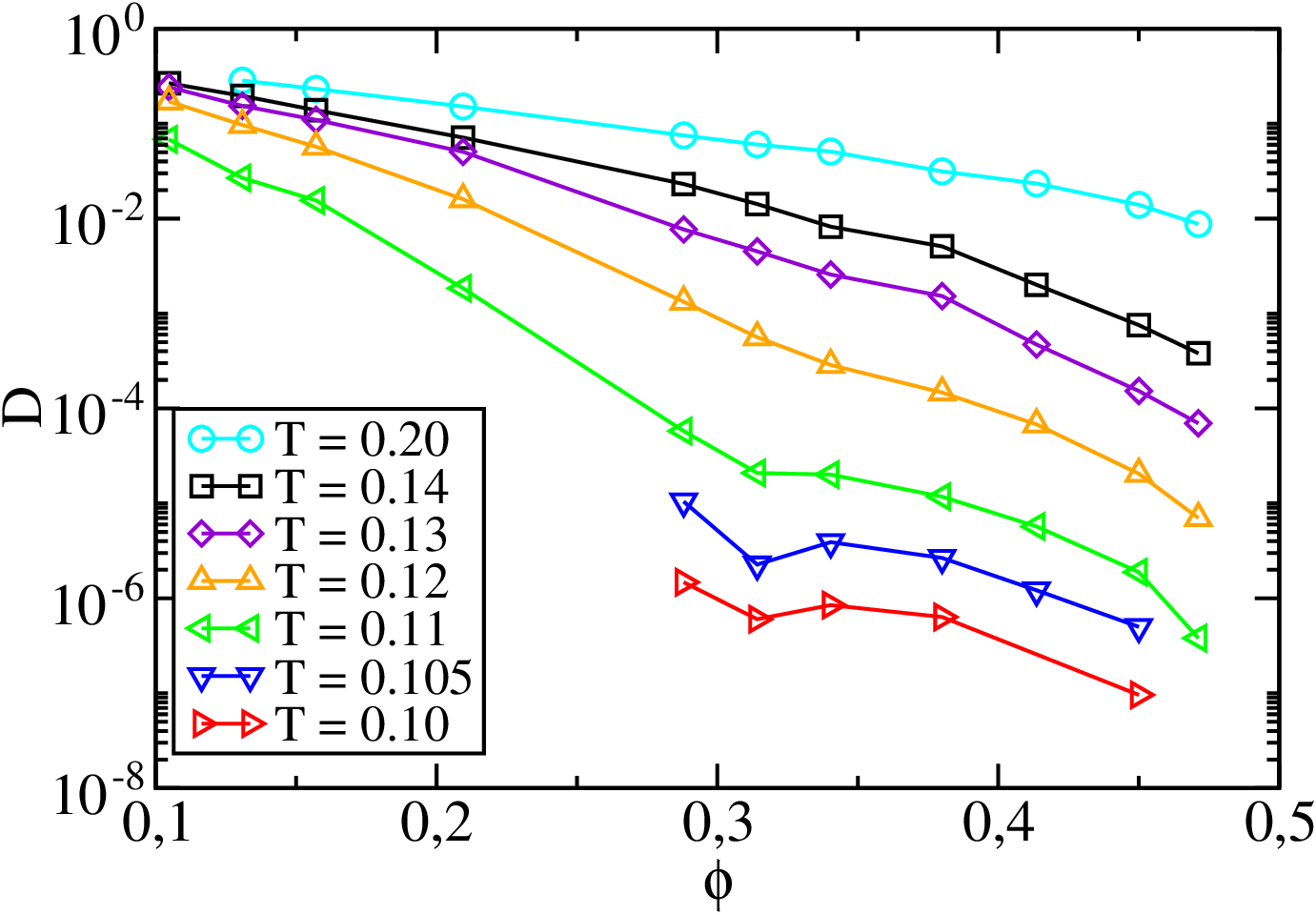}
%\vspace{0.10cm}
\caption{Diffusion coefficient along isotherms. Note the non-monotonic behavior which develops for
$T<0.11$.}
\label{fig:diff}
\end{figure}

\subsection{Isodiffusivity  (and arrest) lines}

\begin{figure}[tbh]
\centering
\vspace{0.10cm}
\includegraphics[width=0.45\textwidth]{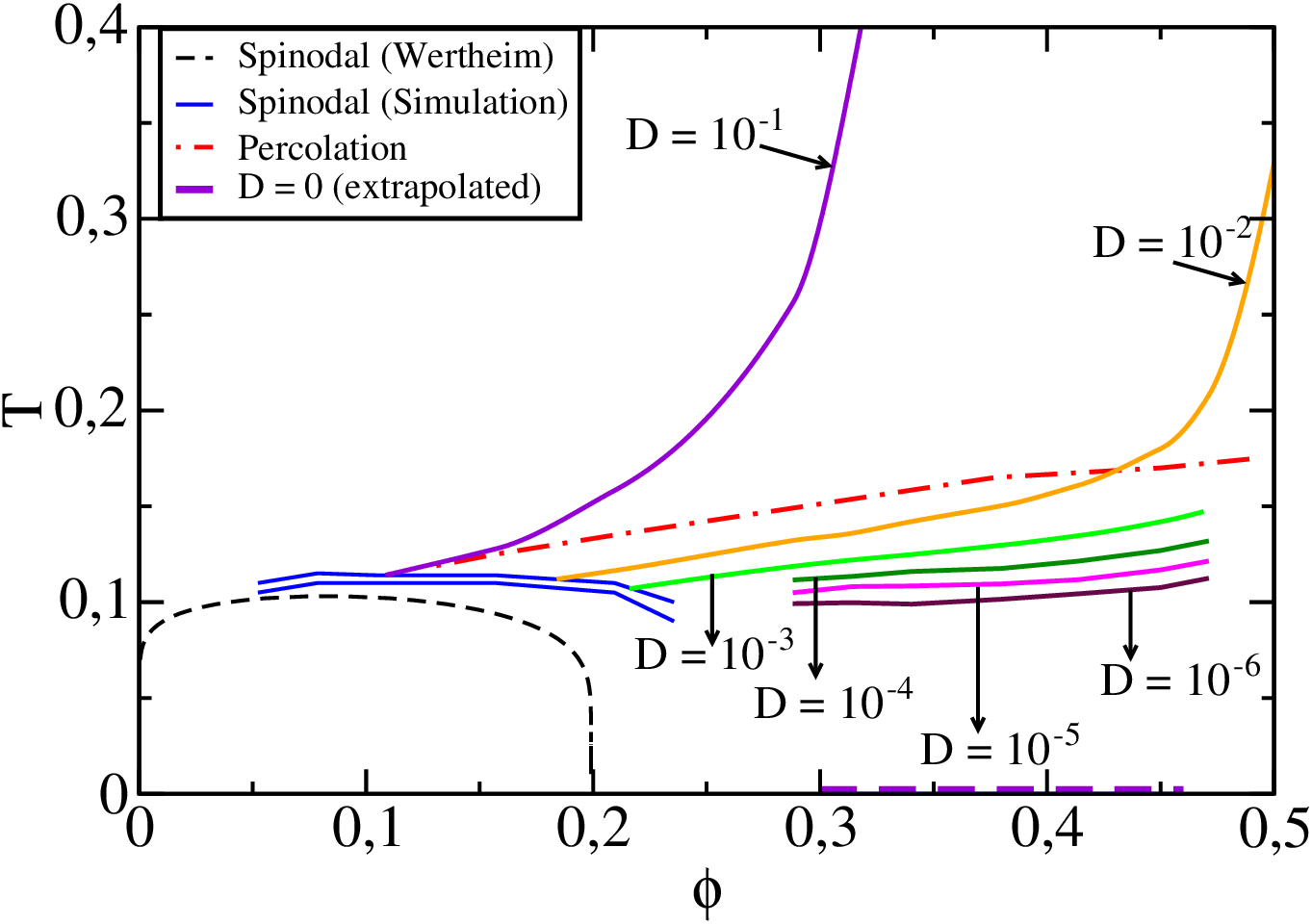}
%\vspace{0.10cm}
\caption{Isodiffusivity lines in the $(T-\phi)$ plane. An excursion of five orders of magnitude in $D$ values is explored. All lines start from the spinodal and end at
infinite $T$ at the corresponding HS location. At small $D$,   lines cannot be continued above $\phi=0.5$ since there the HS interaction is dominant and the system crystallizes. 
Extrapolating along isochores the observed Arrhenius  functional  form suggest an
ideal $D=0$ arrest line at $T=0$.}
\label{fig:isod}
\end{figure}

A global view of the dynamics in the $(T-\phi)$ plane is offered by the
isochronic lines, i.e. the locus of state points with the same
characteristic time\cite{Tol01a}. In the present case we focus on the isodiffusivity
lines. The shape of the isodiffusivity lines, extrapolated to 
$D  \rightarrow 0$  provides a useful indication of the shape of the glass transition line\cite{Fof02a,Zac02a,Sci04a}.  Fig.~\ref{fig:isod} shows the
isodiffusivity lines for several different values of $D$, separated each other by one order of magnitude. The slowest isodiffusivity lines are only weakly $T$ dependent at low $\phi$. For small values of $D$,
iso-diffusivity lines start from the right side of the spinodal, confirming that slow dynamics is only possible for states with $\phi>\phi_c$.   At large $\phi$ the isodiffusivity lines bend and become parallel to the $T$ axis, signaling the  cross-over to the hard-sphere case. 
Extrapolating to zero the $T$ (or $\phi$) -dependence of $D$ it is possible
to provide estimates of the dynamic arrest line.  In the present model, the
low $T$-dependence of $D$  along isochores is well modeled by the
Arrhenius law and hence technically arrest is expected at $T=0$. 
The shape of the iso-diffusivity lines  suggests that the vertical repulsive glass line (controlled by excluded volume effects) starting at high $T$ from the HS glass packing fraction meets at a well defined $\phi$ the $T=0$ bond glass line.

The shape of the PMW isodiffusivity lines is very similar to the 
short-range square well case, for which a flat $T$-independent 
"attractive" glass line crosses (discontinuously) into a perpendicular
$\phi$ independent "repulsive" glass line\cite{Daw00a,Zac02a}. 
Differently from the SW case, in the PMW the equivalent of the attractive glass line extends to much smaller $\phi$ values, since the reduced valency has effectively reduced the space in which phase separation is observed\cite{Zac05a}.
It is also worth pointing that the shape of the isodiffusivity lines at low $\phi$ is similar to the shape of the percolation line. As in all previously studied models\cite{Zac02a,Zac05a}, crossing the percolation line does not coincide with dynamics arrest, since the bond lifetime is sufficiently short that each particle is able to break and reform its bonds.

\subsection{$D$ vs. $E-E_{gs}$}
\begin{figure}[tbh]
\centering
\vspace{0.10cm}
\includegraphics[width=0.45\textwidth]{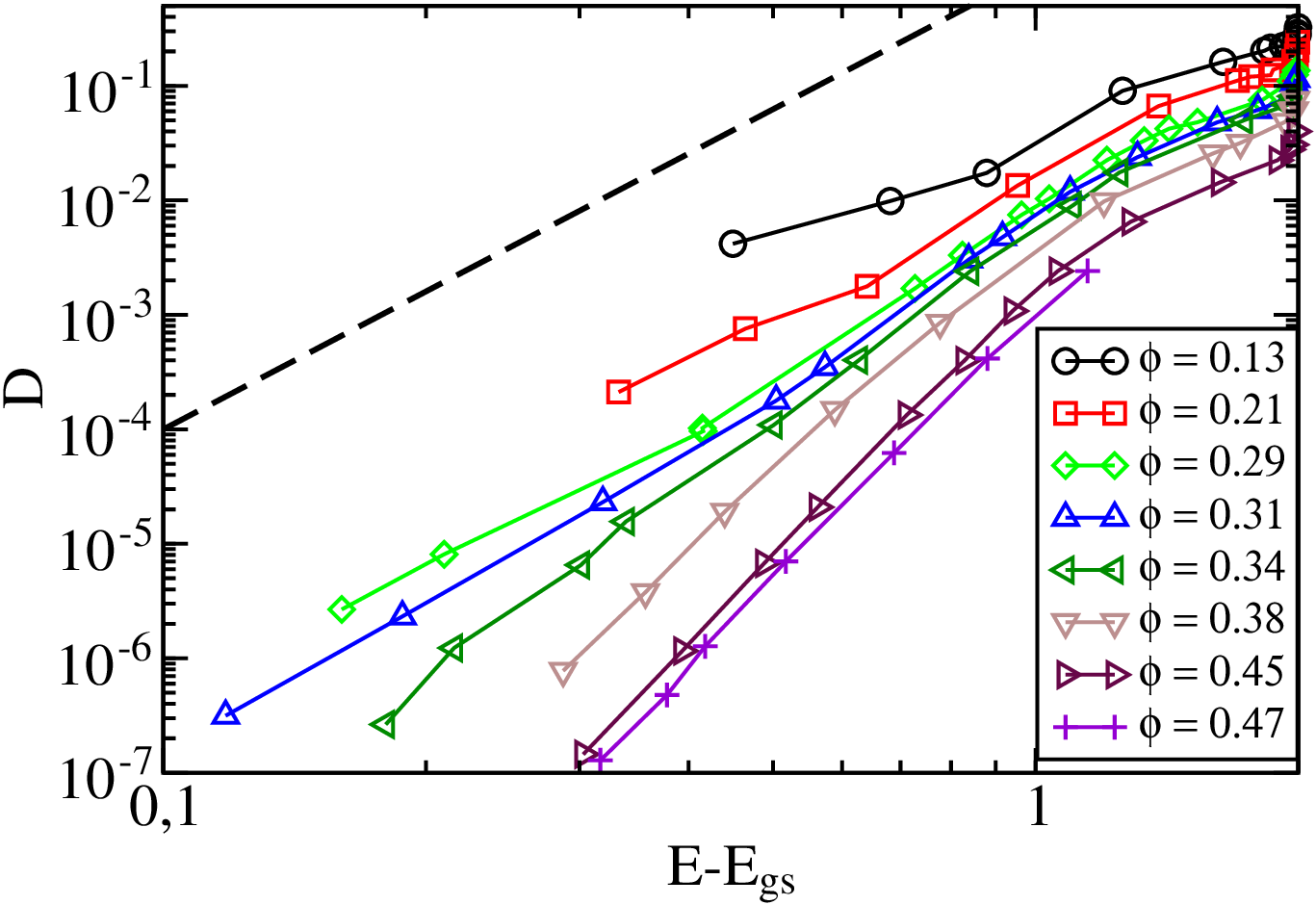}
%\includegraphics[width=0.5\textwidth]{DvsEdiPHI.eps}
%\vspace{0.10cm}
\caption{Diffusion Coefficient vs $E-E_{gs}$ for different $\phi$ values. The dashed line
is a power-law with exponent four.}
\label{fig:DvsE}
\end{figure}

At the optimal network density, 
the low $T$ behavior of both $D$ and $E-E_{gs}$ (which, as discussed above, is also a measure of the number of broken bonds) is Arrhenius. This suggests to look more carefully in the relation between the activation energy of the two processes. One possibility is offered by a parametric plot  of  $D$ vs $E-E_{gs}$ in log-log scale, so that the slope of the straight line provides the ratio of the two activation energies. Such a plot is shown in Fig.~\ref{fig:DvsE}.  We find the remarkable results that close to the optimal network $\phi$, the slope of the curve has exponent four, i.e.
$D \sim (1-p_b)^4$,  where $p_b$ is the  probability that one of the four bonds is formed (and hence $1-p_b$ is the probability that one of the four possible bonds is broken), suggesting that the elementary diffusive process requires the breaking of  four bonds.  A  functional law for diffusion in a tetrahedral model of this type was  proposed by Teixera\cite{Tei90a}  to interpret the $T$ dependence of $D$ in water in the context of the percolation model developed in Ref.~\cite{Sta80b}. A similar dependence has been recently reported for a model of gel-forming four-armed DNA dendrimers\cite{Sta05a}.

\subsection{$D$ - MD vs. MC}
\begin{figure}[tbh]
\centering
\vspace{0.10cm}
\includegraphics[width=0.45\textwidth]{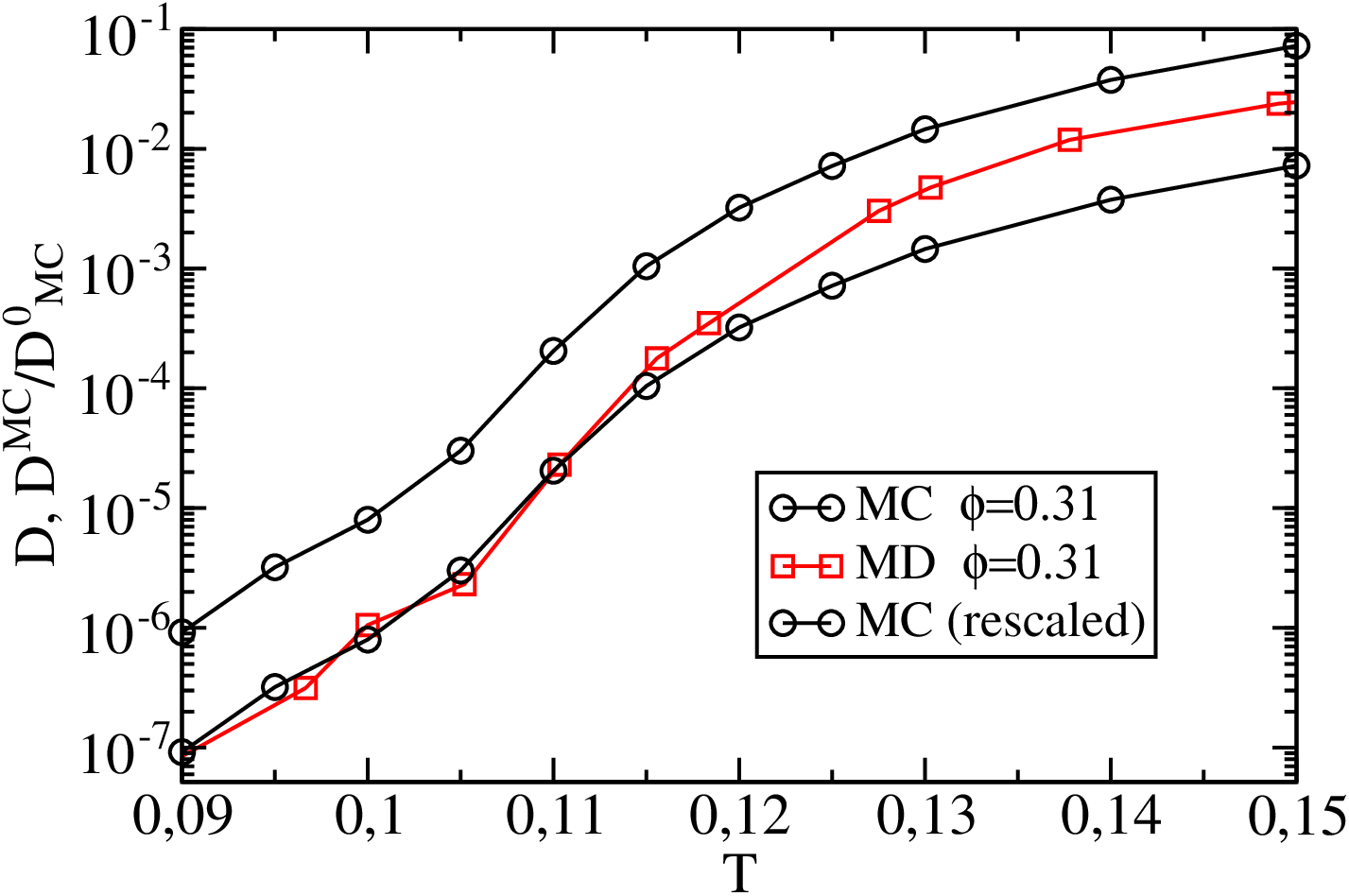}
\includegraphics[width=0.45\textwidth]{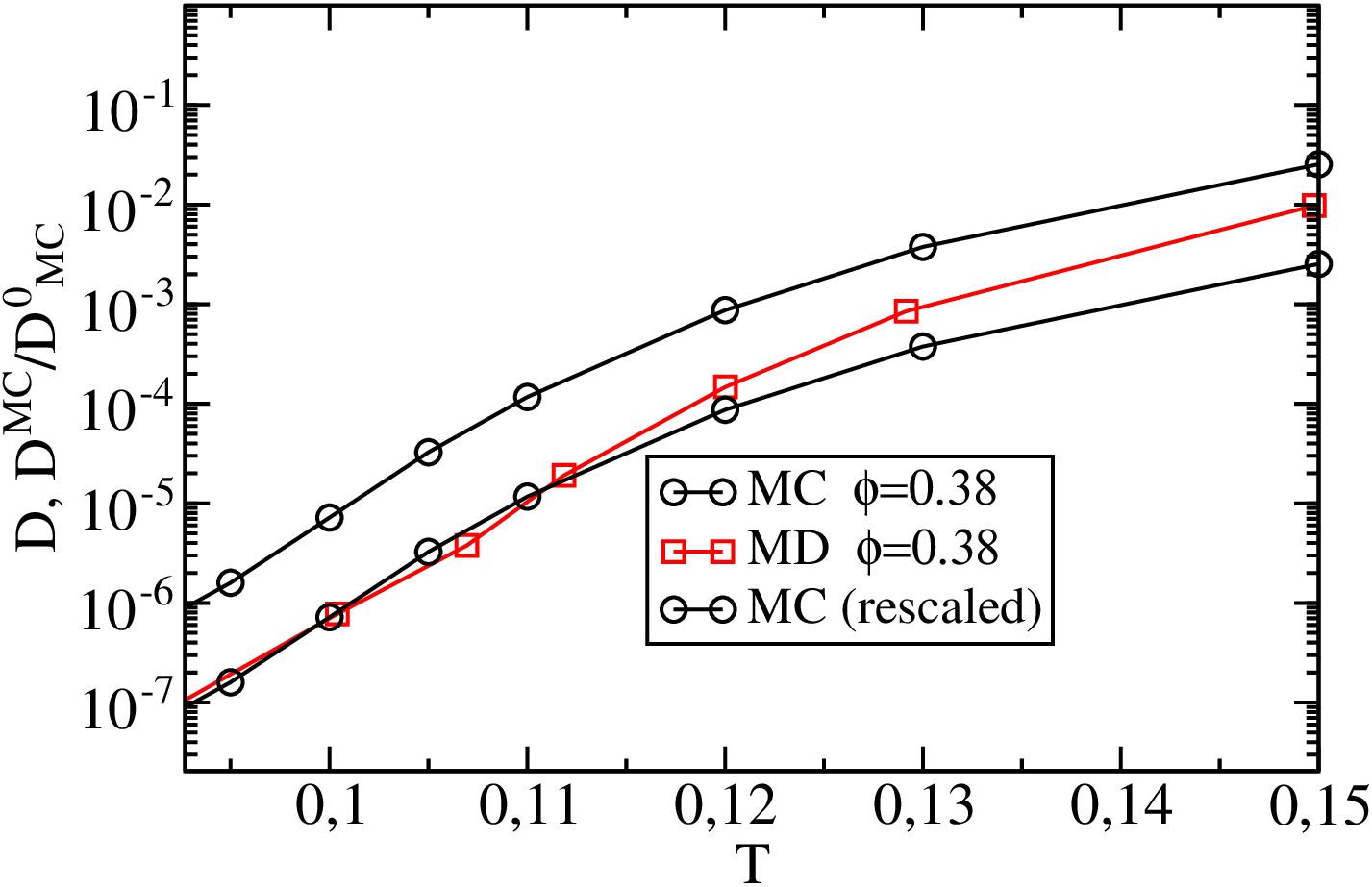}
\includegraphics[width=0.45\textwidth]{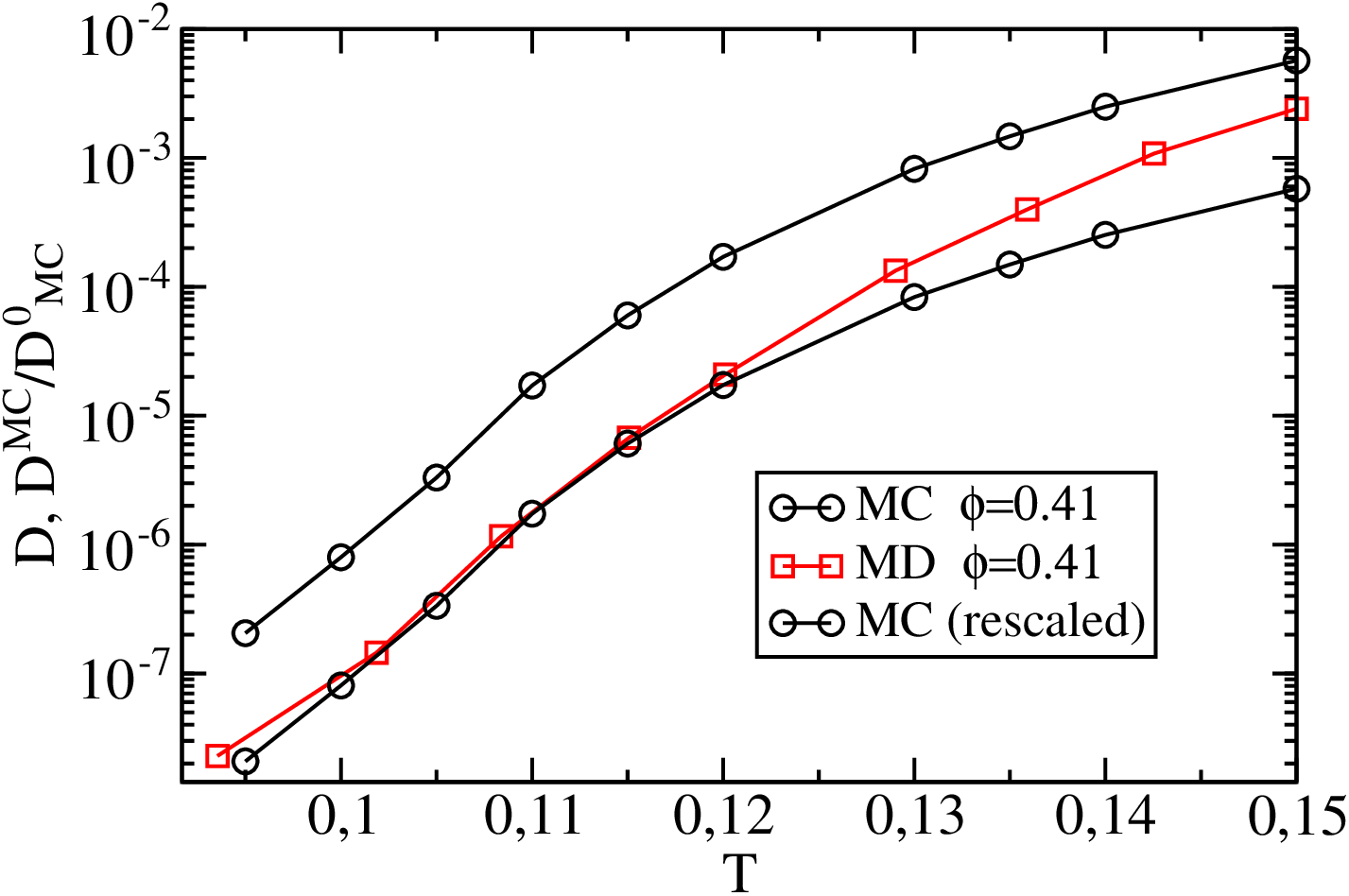}
%\includegraphics[width=0.5\textwidth]{DvsEdiPHI.eps}
%\vspace{0.10cm}
\caption{Comparison between the $MD$ and $MC$ diffusion coefficient at three different $\phi$ values. The MC data are also shown multiplied  (by a common factor $0.1$) to better visualize the low $T$ overlap.}
\label{fig:DMCMD}
\end{figure}

All dynamic data presented above refer to event-driven newtonian dynamics. Indeed,
Monte Carlo simulations intrinsically miss dynamic informations being based, in their simpler formulations,  on random displacements of the individual particles. Still, if the random displacement
in the trial move is small as compared to the particle size the sequence of MC steps can be
considered a possible trajectory in configuration space.  When this is the case, the number of MC-steps (each step being defined as an attempted move per each particle) plays the role of time in  the evolution of the configurations  in configuration space.  In the absence of interactions, a particle evolved according to
the MC scheme diffuses with a  bare diffusion coefficient $D_{MC}^0$ fixed by the variance  $\delta_{MC}^2$ of the chosen random displacement along each direction (in our calculations we have used an uniform distribution of displacements with a variance of $\delta_{MC}^2=(0.1)^2/12$, corresponding to  $D_{MC}^0=  3 \delta_{MC}^2/6$ in units of $\sigma^2$/MC-step). If needed,  $D_{MC}^0$ provides a mean to associate a physical time to the MC-step. 
At low $T$, when slow dynamic processes set in (favored by bonding or by packing), it is expected that
the microscopic dynamics becomes irrelevant (except for a trivial scaling of time). The escape from the
cage created by neighboring particles is indeed a much rare  event as compared to the
rattling of the particles in the cage. Under these conditions, the  slow dynamic processes  become independent on the microscopic dynamics, and hence Newtonian, Brownian and MC  show the same trends.  Fig.\ref{fig:DMCMD}
shows that  this is the case for three $\phi$ values. In all cases, at low $T$, 
the $T$ dependence of $D^{MC}$ and $D$ is identical. Moreover, the scaling factor
between $MC$ and $MD$ dynamics is independent of $\phi$, suggesting that at low $T$, with the chosen units,  the relation $D^{MC}/D_{MC}^0= \xi D$ holds. From comparing MC and MD
data we find that the proportionality constant $\xi \approx 10 $  and shows no state-point dependence. 
To confirm that  caging is fundamental to observe independence of the slow-dynamics from the microscopic one, we look at the shape of $<r^2(t)>$ (Fig.~\ref{fig:msd}), finding  that at the  $T$  at which  MC and MD dynamics start to coincide a significant caging is present. 

\begin{figure}[tbh]
\centering 
\vspace{0.10cm}
\includegraphics[width=0.45\textwidth]{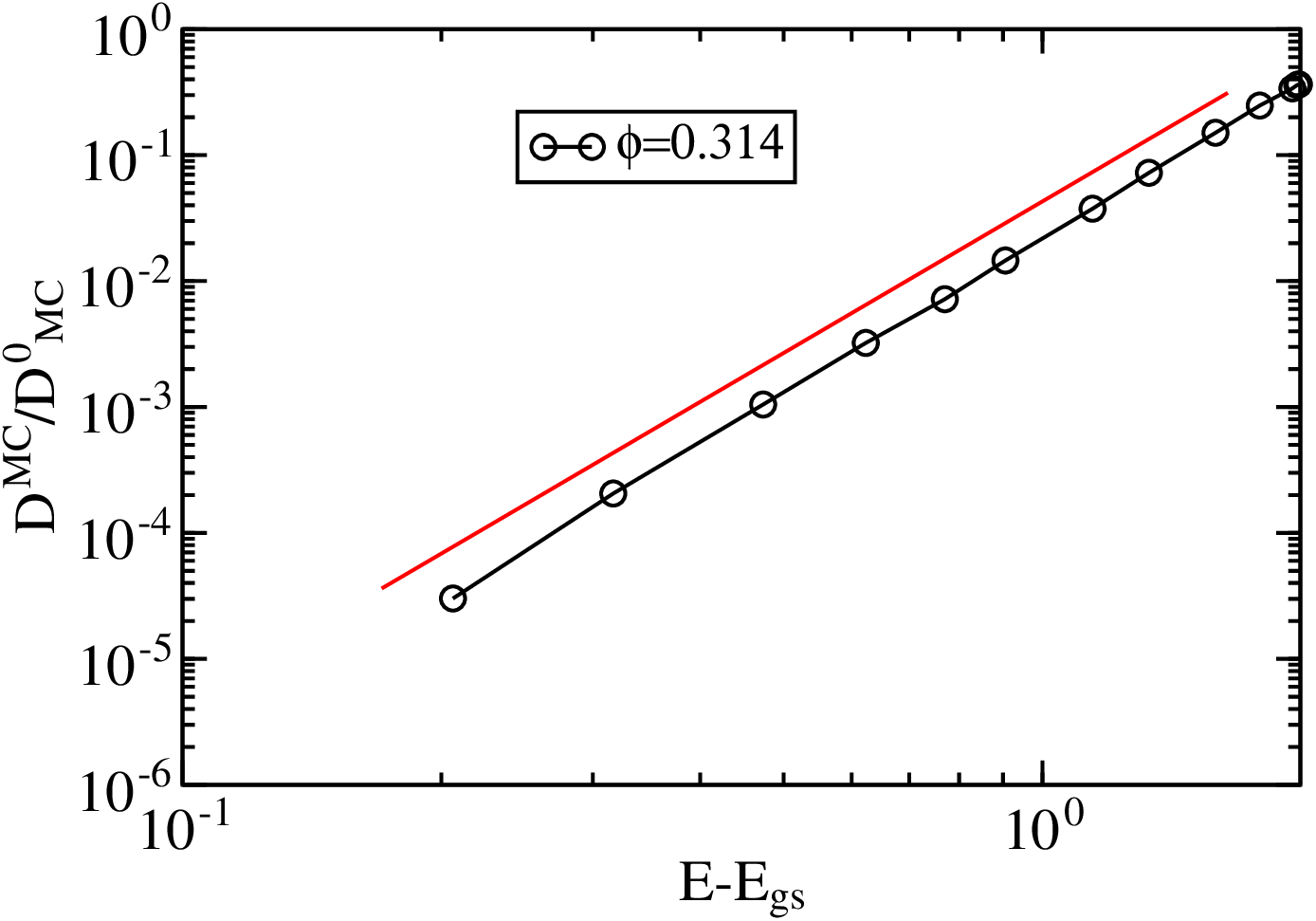}
\caption{Relation between $D^{MC}$, normalized by the bare MD diffusion constant $D^0_{MC}$
and $E-E_{gs}$ for MC dynamics. Note that the
MC data follow over more than five orders of magnitude a simple fourth-power law (full red line). }
\label{fig:mcdvse}
\end{figure}

Since the microscopic time of the MC dynamics is not affected by temperature (being always fixed by the variance of the random displacements) it is interesting to consider the relation between
$D$ and  $E-E_{gs}$ also for $D^{MC}$, shown in Fig.~\ref{fig:mcdvse} at the
optimal network density $\phi=\phi=0.314$.  Again, the slope of the curve has exponent four, but
compared to the MD case,  the region of validity of the power-law covers
the entire range of $T$ studied, from very high $T$ (where
the number of bonds is negligible) down to the lowest equilibrated temperature, covering more than 4 order of magnitude.  The validity of the relation $D \sim (1-p_b)^4$ extends up to high $T$, when
the system is well above percolation and there is no evidence of a tetrahedral network (as shown in the structural data reported in Fig.~\ref{fig:sqoo} and \ref{fig:groo}). 
The extended validity of the power-law, with an exponent exactly equal to the valence of the model is highly suggestive and, in principle, very important for theoretical considerations, since it appears to cover either the region of temperature where liquid dynamics is observed,  either the low $T$ states where signatures of slow dynamics (see Fig.\ref{fig:msd}) are very well developed. The limit of
validity of this finding needs to be carefully checked in other  primitive models with different valence and with more realistic models of network forming liquids.

\section{Conclusions}

Results presented in this manuscript cover several apparently distinct fields. 

To start with, results presented here can be discussed in relation to the dynamic and thermodynamic properties of water.  We have shown that
the thermodynamic of the PMW includes, beside the compressibility anomalies reported before\cite{Kol87a}, also  density anomalies (at much lower $T$). 
The source of the density anomalies is shown to be associated to the
establishment of the bond network in the tetrahedral geometry. 
On cooling (along isochores) the energetic driving force which favors the formation of the bond, due to geometric constraints associated to the formation of the open tetrahedral structure, forces the pressure to increase and hence generating a density maximum state point.   The simplicity of the
PMW allows us also to clearly detect an optimal network density, at which 
the ground state of the system (i.e. the state in which each particle is involved in four bonds) can be closely approached. At this $\phi$ the 
$T$-dependence of the potential energy is the most pronounced, generating
a minimum in the isothermal $\phi$ dependence.  The presence of a minimum in $E(\phi)|_T$ is highly suggestive since it indicates\cite{Sci97a} the  possibility of a liquid-liquid phase separation at $T$ lower than the one we have been able to equilibrate. We have also shown that at this optimal $\phi$, low $T$ dynamics slows down with the fourth power of the probability of broken bonds, i.e. the dominant component to dynamics arises from single particle motions, and specifically of the particles which happen to have all four bonds broken at the same time. 
We have also shown that, like in real water, diffusion anomalies are
observed.  At low $T$, the decrease of the diffusivity on increasing $\phi$ 
is reversed once the optimal network density is reached. For higher $\phi$, the progressive destruction of the bond network due to the increased packing fastens the dynamics. For even higher $\phi$, $D(\phi)$ resumes its standard decreasing behavior associated to the approach of the
excluded volume glass transition.   Diffusion and density anomalies 
in the PMW models are thus strongly related, similarly to what has been observed in more realistic models for water\cite{Err01aNature}.
The simplicity of the model is crucial in clarifying these aspects since
the hard-core and square well interactions guarantee the absence of
volumetric effects related to the $T$-dependence of the vibrational amplitudes. 

A second interesting aspect of the presented  results concerns the
dynamics in network forming systems. The present study provides a
complete characterization of the dynamics in the entire $(\phi-T)$ plane,
from the smallest possible $\phi$ liquid state points up to the close packed state.  From the reported data,  the relative role of the energy and of the packing in controlling the dynamics stands up clearly. The isodiffusivity lines are essentially parallel to the $\phi$-axis (i.e. $T$ controlled) in the 
network low $\phi$ region and are essentially parallel to the $T$-axis (i.e.
$\phi$ controlled) at larger $\phi$. Interesting enough, along isochores, 
low $T$ dynamics follows and Arrhenius law, the landmark of
strong-glass forming behavior\cite{fragile,Deb01aNature}. The Arrhenius law is foreseen by a $T$
region where dynamics has a strong $T$ dependence, compatible with a
power-law dependence.  In this power-law region the first signatures of caging in the mean square displacement are observed. 
Similar changes in the dynamics have been observed in previous studies  of silica\cite{Hor99a,Hor01aPRE,Voi01a}, water\cite{Xul05aPNAS} and silicon\cite{Sas03aNatMat}.
In particular, for the case of silica and water, it has been suggested that
the region where dynamics start to feel the presence of energetic cages can be interpreted in terms of mode coupling theory\cite{Sci96b,Sci01aPRL,Sta99aPRL,Fab99a,Kob01aJNCS,Sci00a,Hor99a,Hor01aPRE}. 

Dynamics at the optimal network $\phi$  is particularly suggestive.
Although in the present model, slowing down of the dynamics prevents equilibration of the supercooled liquid  to very low $T$, at the lowest $T$ simulations the average number of bonds has gone up to 3.8 per particle.  In this respect, further structural and dynamic changes are hard to 
foresee.  This suggests that the Arrhenius behavior is retained down to
$T=0$. Such speculation is reinforced by the numerical values  of the activation energy of $D$ which is found to be $\approx 4 u_0$, i.e. corresponding to the breaking of four bonds.  This suggests that in network liquids, the limited valency  imposed by the directional forces fixes a well defined energy  of the local configuration and a discrete change of it which is reflected in the Arrhenius behavior.  The presence of a limited valency and a well defined bond energy scale appears to be the key ingredient of the strong liquids behavior\cite{Mor05a}. It is also worth to explore in future works the possibility
that the optimal network density plays, in studies of one component systems, the same
role as the reversibility window\cite{reversibility} in bulk alloy glasses.  Connections
with the concept of self-organization in network glasses\cite{naumis} should also be pursued. 

A further aspect of this work concerns the relative location between the
liquid-gas spinodal and the kinetic arrest lines, whose shape is
inferred by the study of the isodiffusivity lines.  As in the short range SW model\cite{Zaccapri,Fof05aPRL}, the kinetic arrest lines ends in the right side of the
spinodal, i.e. in the liquid phase. But differently from the SW case,
the limited valency has shifted the right side of the spinodal to 
very small $\phi$ values, $\phi \lesssim 0.25$. Indeed, the limited valency effectively disfavors condensation of the liquid phase reducing the driving force for phase separation and making it possible  to generate low packing fraction arrested states in the absence of phase separation, i.e homogeneous  single phase stable in equilibrium, at
low $T$\cite{Sci04bCPC}.  The possibility to access low $T$ homogeneous supercooled states  for $\phi>0.25$ characterized by a glassy dynamics, 
driven by the bonding energy as opposed to packing,  confirms the findings of the zero-th order model with limited valency reported in Ref.~\cite{Zac05a}.
The  absence of  geometric correlation between the bonding sites, the key ingredient of the maximum valency model\cite{Zac05a} is thus not crucial for the stabilization of the network. The role of the geometric constraint appears to be  the reduction in the range of $\phi$ values where the fully bonded disordered  state can be reached.  Two different arrest mechanisms characterize the
dynamics of network systems. Arrest due to the formation of energetic cages, with an arrest line which runs almost parallel to the $\phi$ axis, and
arrest due to excluded volume effects, with an arrest line
parallel to the $T$ axis.    These two lines are reminiscent of the
attractive and repulsive glass lines observed in short-range 
attractive colloids\cite{Fab99a,Ber99a,Daw00a,Zac02a,Sci02a}. 
%Differently from the spherical potential case, where the attractive glass line is confined to a small region of $\phi$-values, here the shrinking of the spinodal makes it possible to observe a fully developed arrest line at small $\phi$. 
Connecting the results presented in this article with previous studies 
of network forming liquids\cite{Hor99a, Sci96b}, it is tempting to speculate that
mode-coupling theory predicts satisfactory  the shape
in the $(\phi-T)$ plane of the dynamic arrest lines. Still, while in the region where  excluded volume controls caging  the relative error in the location of the glass line is limited , in the case in which   bonding mechanism is dominant in generating arrest, the location of the MCT line can be significantly distant from the actual dynamic arrest line (technically located at $T=0$, being dynamics Arrhenius), due to the
role of activated bond-breaking processes which offer a faster channel for the decay of the correlations.  The evaluation of the MCT lines for the
present model, in principle feasible within the site-site approach developed by Chong and Goetze\cite{Cho98a,Cho02b} or within the molecular approach developed by Schilling\cite{Sch97a,Fab99b}  can help clarifying this issue.      

The possibility of an intersection between the excluded volume arrest-line (starting at high $T$ from the HS glass packing fraction ) and the bond-controlled $T=0$ arrested line     is particularly suggestive.  The shape of the iso-diffusivity lines supports the
possibility that  the vertical repulsive glass line meets at
a well defined $\phi$ the $T=0$ bond-controlled  glass line.   
If this scenario is correct and general, one would conclude that the fragile and strong kinetic behavior is intimately connected to the dominant mechanism of arrest (fragile for excluded volume and strong for bonding) and, more interestingly, that strong behavior can only be observed when the interaction potential is such that less than six neighbors are present (i.e. in network forming systems). Indeed, only under these circumstances  the suppression of the liquid-gas phase separation makes it possible to approach the $T=0$ bond-controlled glass line.
 
An additional comment concerns the relation between gel  and glass arrest states.
Results reported in this article confirm, one more, that in this class of models the geometric percolation line does not have any impact on the dynamic arrest, since at percolation the lifetime of the bond is still rather small. Only when the system is well inside the percolation region, the bond lifetime has slowed down significantly to affect all measurements of global connectivity with an intrinsic time scale shorter than the bond lifetime  (as for example finite frequency shear viscosity).  Indeed, already long time ago it was noted  for the case of water\cite{Sta80b} that bond percolation is irrelevant to any thermodynamic or dynamic anomaly.  More sophisticated models, incorporating bond-cooperativity or significant entropy contributions to bonding (as the case of polymeric gels) may reduce the differences between dynamic arrest states  and percolation\cite{Sta05a}.

Despite the difference between percolation and arrest lines, if one consider the present model as a system of colloidal particles with sticky interactions, one would be led to call the arrested state at $0.3 \lesssim \phi \lesssim 0.5$ a gel, led by the fact that the arrested state has a low $\phi$ open connected structure.  Similarly, if one consider the PMW as a model for a network liquid, one would be led to name the same arrested state a network glass.   While we cannot offer any resolution to this paradox with the present set of data, future work focusing on the shape of the  wavevector dependence correlation functions and the resulting non ergodicity parameters can help clarifying this issue and confirm/dispute the hypothesis on the differences between gels and glasses recently proposed\cite{Ber99a,Zac05a,Fof05bJCP}.  At the present time, we can only call attention on the fact that a continuous change from energetic cages to excluded volume cages takes place on increasing $\phi$. 
%Simultaneously, dynamics changes its character from Arrhenius towards a  more super-Arrhenius dependence.
%, as compared to polymeric gels,  the bond formation  is essentially controlled by energy      

A final comment refers to the propensity of the system to form  disordered arrested states. Despite the  relevant amount of supercooling\cite{Veg98a}, in all studied state points where a network structure is present, we have not observed any sign of crystallization.  The kinetic suppression of the crystallization phenomenon can be traced to the similar energy characterizing the crystal state and the fully bonded disordered state, vanishing the energetic driving force toward crystallization. The observed propensity to form gel states as opposed to crystalline states manifested by the studied model  (which can be seen also as a model for short-range sticky colloid particles as well as globular proteins with aeolotopic interactions\cite{Lom99a}) may well explain the difficulty of crystallizing some class of proteins.  It also warn us about the relevance of the dynamic arrest phenomenon in the future attempts to build a colloidal diamond photonic crystal, made of particles with short-ranged patchy interactions.

\section{Acknowledgements}
We thank E. Zaccarelli. We acknowledge support from MIUR-FIRB and
CRTN-CT-2003-504712.

%\bibliographystyle{./apsrev}
%\bibliography{./fileunico,./altri}

\section{Appendix: An event-driven algorithm for hard spheres with patches.}
\label{appendicecris}

In an event driven (ED) algorithm,  events such as 
times of collisions between particles and cell crossing have to be taken into account.
All these events have to be ordered. Code must be written in such a way that locating the  
next event  and insertion/deletion of new events have to be performed efficiently.
In literature, several ED algorithms for simulating 
hard-sphere systems exist and several propositions on how to handle such events efficiently have been reported. 
One elegant approach, proposed twenty years ago 
by Rapaport \cite{Rapaport},  arranges events into an ordered 
binary tree (calendar of events)
so that insertion, deletion and retrieving of events can be done with an efficiency 
$O(\log N)$, $O(1)$ and $O(\log N)$  respectively, 
where $N$ is the number of events in the calendar. 
We adopted this solution to handle the events calendar in our simulation, adding only
a redefinition of event time in order to avoid round-off problems which are found when
extremely long simulation runs are performed. 

\subsection{Motion of rigid bodies}
The orientation of a rigid body can be conveniently represented by the $3$ column 
eigenvectors ${\bf u}_i$ (with $i=1,2,3$) of the 
inertia tensor expressed in the laboratory reference system.
These vectors form an orthogonal set and can be arranged in a matrix  ${\bf R}$, 
i.e.
\begin{equation}
{\bf R} =  {}^t ( {\bf u}_0\  {\bf u}_1\  {\bf u}_2 )
\end{equation}
where ${}^t A$ indicates the transpose of the matrix $A$.
This matrix is such that if ${\bf x}$ are the coordinates of the laboratory reference system
and ${\bf x'}$ are the coordinates of the rigid body reference system, it turns out that:
\begin{equation}
{\bf x}' = {\bf R} {\bf x}
\end{equation}
In what follows, we assume that the three eigenvalues of the inertia tensor are all equal to $I$.
Naming ${\bf w}=(w_x,w_y,w_z)$ the angular velocity of a free rigid body,  
the matrix ${\bf \Omega}$ is defined as
\begin{equation}
{\bf \Omega} = \begin{pmatrix}0 & -w_z & w_y\cr w_z & 0 & -w_x\cr -w_y & w_x & 0\end{pmatrix}
\label{Eq:omegmat}
\end{equation}
Knowing the orientation at time $t=0$, the orientation ${\bf R}(t)$  at time $t$ is:
\cite{LandauMec,Goldstein}:
\begin{equation}
{\bf R}(t) = {\bf R}(0) ({\bf I} + {\bf M})
\label{Eq:Rt}
\end{equation}
where ${\bf M}$ is the following matrix:
\begin{equation}
{\bf M} = - \frac{\sin(wt)}{w} {\bf \Omega} + \frac{1-\cos(wt)}{w^2} {\bf \Omega}^2
\label{Eq:Mmat}
\end{equation}
and $w = \|{\bf w}\|$.
Note that if $w=0$ then ${\bf R}(t) = {\bf R}(0)$.
To derive Eq. \ref{Eq:Rt}, consider that:
\begin{eqnarray}
{}^t R(t) &=& ({\bf u}_1(t)\  {\bf u}_2(t)\  {\bf u}_3(t) )   \\ \nonumber
&=& ({}^t ({\bf I} + {\bf M}) {\bf u}_1\  {}^t ({\bf I} + {\bf M}) {\bf u}_2\ {}^t ({\bf I} + {\bf M}) {\bf u}_3\ )
\end{eqnarray}
where we remember that ${\bf u}_i$ are column vectors.
Hence, if ${\bf w} = w {\bf\hat n}$, we have after some algebra:
\begin{equation}
{\bf u}_i (t) =  {\bf u}_i \cdot {\bf\hat n}\; {\bf\hat n} + \cos(wt) ( {\bf u}_i -  
{\bf\hat n}\cdot {\bf u}_i\> {\bf\hat n} ) + \sin(wt) \; {\bf\hat n}\times{\bf u}_i   
\end{equation}
that is the so-called {\it Rodriguez's formula} or {\it Rotation formula}, i.e.
a rotation of an angle $wt$ around the axis ${\hat n}$.
To conclude if one has to update position and orientation of a rigid body,
that is freely moving,
this can be accomplished doing:
\begin{subequations}
\label{Eq:surf}
\begin{equation}
{\bf x}(t) = {\bf x}(0) + {\bf v} t 
\label{Eq:surfa}
\end{equation}
\begin{equation}
{\bf R}(t) = {\bf R}(0) ({\bf I} + {\bf M})
\label{Eq:surfb}
\end{equation}
\end{subequations}
where ${\bf x}(t)$ is the position of the center of mass of the rigid body at time $t$ and
$\bf v$ is its velocity.
\subsection{Hard-Sphere with interacting patches}
In the present model, each particle is modeled as an hard sphere with $n$ spherical patches arranged in fixed site locations. In the present case, the site-site interaction is a SW potential, 

\begin{equation}
u_{SW}=
\begin{cases}
-u_0 & \hbox{if}\ \  r < \delta \cr
0 & \hbox{otherwise}\cr
\end{cases}
\end{equation}
where $\delta$  and $u_0$ are the width and the   depth of the SW .  For the following discussion,
the SW interaction can be visualized as a sphere of diameter $\delta$ centered on the site location. Similarly,
one can visualize the particle as a rigid body composed by the hard-sphere joined to the spheres
located on the sites.   In what follows, we identify a particle with the resulting surface.
Defining distance  $d_{AB}$ between two particles $A$ and $B$ as the shortest line connecting two
points on distinct particles, i.e. 
\begin{equation}
d_{AB} = \min_{i_A,i_B}{d_{i_A i_B}}
\label{Eq:distance}
\end{equation}
where $i_{A},i_{B}\in\{0,\dots n\}$ and $0$ labels the hard sphere, $1\dots n$ label the $n$ spherical patches and $d_{i_A i_B}$ is the distance between the two spherical patches $i_A$ and $i_B$.
\subsection{Prediction of time-of-collision}
\subsubsection{Finding the contact time}
We separate the collisions between two particles in the hard-sphere part of the potential and the 
site-site interaction part. The time of collision $t_{hs}$ between the hard sphere cores  can be evaluated as usual \cite{Rapaport} . The smallest time of collision among all $n^2$  spherical patch pairs is $t_{st}$. 
Time of collision of the two particles is
\begin{equation}
t_c = \min\{t_{hs},t_{st}\}
\end{equation}
To find the time-of-collision of two interacting patches we assume that it is possible to bracket it.
I.e., we assume  (see further subsections) that the time of collision 
$t_{st}$ is such that  $ t_1 < t_{st} < t_2 $ where the product $d(t_1) d(t_2) < 0$.  Thus, the ``exact`` time of collision is provided by the root of the following equation:
\begin{equation}
\| r_{i_A}(t) - r_{i_B}(t) \|  = \delta 
\label{Eq:tc}
\end{equation}
where $r_{i_A}$ and $r_{i_B}$ are the two site locations.

\subsubsection{Linked lists and Centroids}
As described in \cite{Rapaport} to speed up a ED molecular dynamics of hard spheres one can 
use linked lists.
For  a system of $N$ identical particles 
inside a cubic box of edge $L$,  we define  the  ``centroid'' \cite{torquato1,torquato2}
as  the smallest sphere that contains the particle (the HS and the spherical patches).   
Linked lists of centroids  may be quite useful to reduce the number of objects to check for possible collisions and in addition they can be used to restrict the time interval within which searching 
for the collision.
We divide the box into $M^3$ cells so that into each cell contains at most 
one centroid.
After that we build the linked lists of these centroids and handle these lists 
as done in a usual ED molecular dynamics of hard spheres \cite{Rapaport}.
This means that whenever an object cross a cell boundary one has to remove such an object from
the cell the particle is coming from and add this object to the cell where it's going to.

Now consider that one has to predict all the possible collisions of a given particle, which is
inside a certain cell $m$.
As for the hard spheres case we take into account only the particles inside the adjacent 
cells (see \cite{Rapaport} for more details) and we predict the times of collisions with these objects.
Consider now two particles $A$ and $B$ at time $t=0$ and their centroids $C_A$ and $C_B$.
Three possible cases arise:
\begin{enumerate}
\item $C_A$ and $C_B$ do not overlap and, evaluating their trajectory no collision between the
two centroids is predicted.  In this case $A$ and $B$ won't collide as well.
\item $C_A$ and $C_B$ do not overlap but they will collide: in this case,  we 
calculate two times $t_1$ and $t_2$, bracketing the possible collision between $A$ and $B$: 
$t_1$ is defined as the time when the two centroids collide and start
overlapping and $t_2$ is the time when  the two spheres have completely crossed each other and do not overlap any longer.
\item $C_A$ and $C_B$ overlap: in this case  $t_1 \equiv 0$ and $t_2$ is defined as the time at which the two centroids stop overlapping.
\end{enumerate}

%Anyway if the simulated objects are too elongated or more generally if the ratio $V / V_{centroids}$ is small,
%where $V$ is the volume of the simulated rigid body and $V_{centroids}$ is the volume of the corresponding
%centroid, 
%(see \ref{Sec:NNL} for the case of very prolate/very oblate HE) 
%the linked lists alone become inadequate to get good performances and one has
%to develop further tricks.
%A feasible and simple possibility is to use well-known nearest neighbour lists NNL 
%in addition to the linked lists developed so far.
%To implement the NNL for a general rigid body one can cover the rigid body with one or more bounding boxes
%and..
\subsubsection{Fine temporal bracketing of the contact time}
Here we show how a refined bracketing of  solution of Eq. \ref{Eq:tc}  can be accomplished.
First of all we give an overestimate of the rate of variation of the distance between two patches
$i_A$ and $i_B$, that is:
\begin{eqnarray}
\dot d_{i_A i_B}(t) &=& \frac{d}{dt} \left ( \| \BS r_{i_A}-\BS r_{i_B} \| - \delta \right )\cr 
% &\le& \| {\bf v}_A - { \bf v}_B + { \bf w}_A \times ({\bf x}_A(t) - {\bf r}_A(t)) - 
%{ \bf w}_B \times ({ \bf x}_B(t) - {\bf r}_B(t))
%\| \cr 
 &\le &  \frac{\BS {\dot r}_{i_Ai_B} \cdot  \BS r_{i_A i_B}}{\| \BS r_{i_A i_B}\|}
\le  \| \BS v_{i_A i_B} \| 
\cr & = & \| \BS V_{AB} + {\bf\omega}_A \times (\BS r_{i_A} - \BS R_A) - {\bf\omega}_B \times (\BS r_{i_B} - \BS R_B) \|\cr
& \le & \| {\bf V}_{AB} \| +  \| {\bf\omega }_A \| L_A +
\| {\bf\omega}_B \| L_B = \dot d^{max}_{i_A i_B}
\label{Eq:distOver}
\end{eqnarray}
where the dot indicates the derivation with respect to time,
${ \bf r}_{i_A}$, ${\bf r}_{i_B}$ are the positions of the two sites with respect to laboratory reference
system,
${\bf v}_{i_Ai_B}$, is the relative velocity of the two sites, 
${\bf V}_{AB}$ is the relative velocity between the centers of mass of the two  particles,  
$\BS R_A$ and $\BS R_B$ are the positions of their centers of mass
and 
\begin{subequations}
\label{Eq:lalb}
\begin{equation}
L_A \ge \max_{{\bf r}'\in A} \{\|{\bf r}'-{\bf R}_A\|\} 
\end{equation}
\begin{equation}
L_B \ge \max_{{\bf r}'\in B} \{ \|{\bf r}'-{\bf R}_B\|\} 
\end{equation}
\end{subequations}

Having calculated an overestimate of $\dot d_{i_A i_B}(t)$ we can evaluate an overestimate of $\dot d_{AB}$
that we call $\dot d_{max}$:
\begin{equation}
\dot d_{max} = \max_{i_A i_B} \{ \dot d_{i_A i_B}^{max} \}
\label{Eq:dmax}
\end{equation}
Using Eq. (\ref{Eq:dmax}) we can easily find an efficient strategy to bracket the solution.
In fact the following algorithm can be used:
\begin{enumerate}
\item Evaluate the distances between all sites that may interact $\{d_{i_Ai_B}(t)\}_{i_Ai_B}$
at time $t$ (starting the first time from $t_1$).
\item Choose a time increment $\Delta t$ as follows:
\begin{equation}	
\Delta t =  
\begin{cases}
\frac{d_{AB}(t)}{\dot d_{max}}, & \hbox{if}\> d_{AB}(t) > \epsilon_f \hbox{;} \cr
\frac{\epsilon_d}{\dot d_{max}}, & \hbox{otherwise.} \cr
\end{cases}
\end{equation}
where the two arbitrary parameters $\epsilon_d $ and $\epsilon_f$  satisfy
$\epsilon_d < \epsilon_f \ll \min\{L_A,L_B\}$. 
\item Evaluate the distances at time $t+\Delta t$.
\item If for at least one pair of patches $(i_A,i_B)$ we find that the product 
      $d_{i_Ai_B}(t+\Delta t) d_{i_A i_B}(t) < 0$  we have bracketed a solution.  We then 
      find the collision times and the collision points solving Eq. (\ref{Eq:tc}) for all pairs. 
      Choose the smallest collision time and terminate.
\item if  pairs of patches are such that $0 < |d_{i_Ai_B}(t+\Delta t)| < \epsilon_d$ 
      and $0 < |d_{i_Ai_B}(t)| < \epsilon_d$,  for each of these pairs
      evaluate the distance $d_{i_Ai_B}(t+\Delta t /2)$, perform a quadratic interpolation of these 
      $3$ points $(t,d_{i_Ai_B}(t))$, $(t+\Delta t/2, 
      d_{i_Ai_B}(t+\Delta t/2)$, $(t+\Delta t, d_{i_Ai_B}(t+\Delta t)$ and
      find if the resulting parabolas have zeros.
      If yes refine the smallest zero solving again Eq. (\ref{Eq:tc}) for all these pairs.

\item Increment time by $\Delta t$, i.e.
\begin{equation}
t\rightarrow t+\Delta t
\end{equation}
\item Go to step 1, if $t<t_2$.
\end{enumerate}

If two particles undergo a ``grazing'' collision, 
i.e. a collision where the modulus of the distance 
stays smaller than $\epsilon_d$ during the collision, the collision could  not be located by the previous 
algorithm due to failure of the quadratic interpolation. 
We have chosen to work with $\epsilon_d \approx 10^{-6}$. For such a choice, we have not
observed grazing collisions during the simulation runs (which would appear in the conservation of the energy). 

The basic algorithm can be improved with simple optimizations. For example one can calculate $\dot d_{i_A i_B}^{max}$ as follows:
\begin{equation}
\dot d_{i_A i_B}^{max} = \| {\bf V}_{AB} \| +  \| {\bf\omega }_A \| L_{i_A} + \| {\bf\omega}_B \| L_{i_B} 
\end{equation}
where 
\begin{subequations}
\label{Eq:lalb2}
\begin{equation}
L_{i_A} = \|{\bf r_{i_A}}'-{\bf R}_A\| 
\end{equation}
\begin{equation}
L_{i_B} = \|{\bf r_{i_B}}'-{\bf R}_B\| 
\end{equation}
\end{subequations}
and if $d_{AB}(t) > \epsilon_f$ the time increment can be evaluated in the following optmized way:
\begin{equation}
\Delta t = \min_{i_A i_B}\{d_{i_Ai_B}(t)/d_{i_A i_B}^{max}\}
\end{equation}
%Finally we note that if, after a collision between two sites $i_A$ and $i_B$ of two particles $A$ and $B$, a bond is created or destroyed, then it's important to  ensure that after the time increment $\Delta t$ the distance between the two patches is negative or positive respectively to not miss a possible collision.
%Furthermore if, at the begin of a searching, the distance between two stick spots is too small (e.g. much smaller than $\epsilon_d$), it's safer to decrement the time a bit (e.g. by $ \epsilon_d / \dot d_{max}$) in order to not miss collisions due to round-off errors.

\subsection{Collision of two particles}
At the collision time,  one has to evaluate the new velocities of centers of mass
and the new angular velocities.
If $\BS x_C$ is the contact point then the velocities after the collision can be evaluated
as follows:
\begin{subequations}
\label{Eq:elcoll}
\begin{equation}
\BS v_A \rightarrow  \BS v_A + m_A^{-1} \Delta p_{AB} \BS{\hat  n}
\label{Eq:elcolla}
\end{equation}
\begin{equation}
\BS v_B \rightarrow  \BS v_B - m_B^{-1} \Delta p_{AB} \BS{\hat  n}
\label{Eq:elcollb}
\end{equation}
\begin{equation}
\BS w_A \rightarrow \BS w_A + \Delta p_{AB} I_A^{-1} (\BS r_A -\BS x_C)\times  \BS{\hat  n}
\label{Eq:elcollc}
\end{equation}
\begin{equation}
\BS w_B \rightarrow \BS w_B - \Delta p_{AB} I_B^{-1} (\BS r_B -\BS x_C)\times  \BS{\hat  n}
\label{Eq:elcolld}
\end{equation}
\end{subequations}
where $\BS{\hat  n}$ is a unit vector perpendicular to both surfaces at the contact point $\BS x_C$,
$I_A$, $I_B$ are the moments of inertia of the two colliding sticky particles, $m_A$, $m_B$ their masses and 
the quantity $\Delta p_{AB}$ depends on the type of the collision.
If we define
\begin{equation}
v_c = ( \BS v_A + \BS w_A\times\BS (\BS x_C - \BS r_A ) 
		- \BS v_B - \BS w_B\times\BS (\BS x_C - \BS r_B ) ) \cdot \BS {\hat n}
%	{m_A^{-1} + m_B^{-1} + I_A^{-1}\|(\BS r_A -\BS x_C)\times  \BS{\hat  n}\| +
%		I_B^{-1}\|(\BS r_B -\BS x_C)\times  \BS{\hat  n}\| } 
\label{Eq:vc}
\end{equation}
If the collision occurring between particles is an hard-core collision, one has: 
\begin{equation}
\Delta p_{AB} = -2 v_c 
\label{Eq:factorHS}
\end{equation}
if the collision occurred between two spherical patches already bonded (i.e. prior the collision the distance
between the two sites is $<\delta$ one has:
\begin{equation}
\Delta p_{AB}=
\begin{cases}
 -2 v_c & \hbox{if}\ \  v_c^2 < 2 u_0 /M_{red} \cr
 - v_c +  \sqrt{v_c^2 - 2 u_0/M_{red}}& \hbox{otherwise}
\label{Eq:factorHSvc}
\end{cases}
\end{equation}
where
\begin{equation}
M_{red}^{-1} = m_A^{-1} + m_B^{-1} + I_A^{-1}\|(\BS r_A -\BS x_C)\times  \BS{\hat  n}\| +
			I_B^{-1}\|(\BS r_B -\BS x_C)\times  \BS{\hat  n}\|
\label{Eq:mred}
\end{equation}
Finally if the collision occurs between two patches that are not bonded   (i.e. the distance
between the two sites is $>\delta$ prior the collision) we have:
\begin{equation}
\Delta p_{AB} = - v_c +  \sqrt{v_c^2 - 2 u_0/M_{red}}
\end{equation}

\section{Appendix: Evaluating the pressure}
\label{pressure}

\subsection{Evaluating the pressure in the ED code}
We define the quantity:
\begin{equation}
\Delta A_{\alpha\beta}(t) = V \int_0^t {\cal P}_{\alpha\beta}(t') dt' 
\label{Eq:intstresspress}
\end{equation}
where  ${\cal P}_{\alpha\beta}$ is the molecular pressure tensor,
\begin{equation}
{\cal P}_{\alpha\beta} V = \sum_{i=1}^{N} M_i V_{i\alpha} V_{i\beta} + \sum_{i=1}^{N}\sum_{j>i}^{N} F_{ij\alpha} (R_{i\beta}-R_{j\beta})
\label{Eq:stresstens}
\end{equation}
The sums in the previous expression involve components (denoted by greek letters), ${\vec V}_i$, ${\vec R}_i$ and
${\vec F}_{ij}$, which are the velocity, the position of center of mass of $i$-th particle 
(mass $M_i$) and the total force acting between particle $i$ and $j$ respectively.
%Lastly we note that in practice $\Delta A(t)$ is averaged over the three possible choices $\alpha\beta=xy,yz,xz$.

In the presence of impulsive forces, 
the stress tensor defined in Eq. (\ref{Eq:stresstens}) is not well
defined,  while  the integral in Eq. (\ref{Eq:intstresspress}) is well defined.
Consider the time interval $(t,t+\Delta t)$. During this interval the quantity 
$\Delta A_{\alpha\beta}(t) $ will vary due to the collisions occurring between particles.
The variation $\delta A(t)$ of $\Delta A(t)$, is:
\begin{displaymath}
\delta A_{\alpha\beta}(t) = \sum_i^{N} M_i V_{i\alpha} V_{i\beta} \delta t + R_{i\alpha} \delta P_{i\beta}      
\end{displaymath}
where $\delta t$ is the time elapsed from last collision occurred in the system and 
$\delta \BS P_{i}$ is the variation of momentum of particle $i$ after the collision, i.e.
\begin{equation}
\delta P_{i\alpha} = \Delta p_{AB}  {\hat  n}_{\alpha}
\end{equation}
where $\Delta p_{AB}$ is the quantity defined in Eq. (\ref{Eq:vc}).

From $\Delta A_{\alpha\beta}(t)$ and $\Delta A_{\alpha\beta}(t+\Delta t)$ 
the average pressure over the interval $\Delta t$ can be evaluated as follows:
\begin{equation}
P = \frac{1}{3 V} \sum_\alpha \frac{\Delta A_{\alpha\alpha}(t+\Delta t) - \Delta A_{\alpha\alpha}(t)}{\Delta t}
\label{Eq:press}
\end{equation}

\subsection{Evaluating $P$ in MC}

In the analysis of MC configurations,  pressure has been calculated as sum of three contributions. A trivial kinetic contribution, equal to $nk_BT$;  a positive HS contribution, which requires the evaluation of the hard sphere radial distribution function $g_{_{HS}}(r)$ at distance $\sigma$ and a negative contribution arising from the SW interaction, which requires both the evaluation of the
$H-LP$ radial distribution function $g_{_{H-LP}}(r)$ at distance $\delta$ as well as  the evaluation of  $<R_{H-LP}(r)>$.   For a pair of $H$ and $LP$ sites whose distance is $r$, 
the quantity $R_{H-LP}$ is defined as  
the projection of the vector  joining the
centers of  the two particles associated to the two sites
along the direction of the unitary vector joining the two sites. The ensemble average $<...>$ is performed  over all pairs of $H$ and $LP$ sites at relative distance $r$ \cite{Kol87a}.

The resulting expression for $P$ is
\begin{eqnarray}
P =n k_B T (1 +   
\frac{4 \phi   } ~~~~~~&~& \\
\nonumber
\left[  g_{_{HS}}(\sigma)-  8 \frac{\delta^2}{\sigma^3}  (1-e^{-1/T}) <R_{H-LP}(\delta)> g_{_{H-LP}}(\delta) \right])
\end{eqnarray}.
%#  beta P HS  ==  2 pi /3 * rho^2 * ghs
%#
%#  beta P a   ==  -2 pi /3 rho^2 * 8 * (1-dexp(-1/T) * rmed * 0.15^2 * gs
%#

\bibliographystyle{./apsrev}
\bibliography{./fileunico,./altri}

\end{document}